\documentclass[11pt,superscriptaddress]{article}
\usepackage{jheppub}
\usepackage{epsfig}
\usepackage{amsmath,amssymb}
\usepackage{amsfonts}
\usepackage{mathrsfs}
\usepackage{units}
\usepackage{slashed}
\usepackage{bm}

\title{\boldmath
Relativistic Correction to $J/\psi$ and $\Upsilon$ Pair Production}

\author[a]{ Yi-Jie Li}
\author[a]{,Guang-Zhi Xu}
\author[b]{,Kui-Yong Liu}
\author[a]{,Yu-Jie Zhang}

\affiliation[a]{School of Physics, Beihang University,
 Beijing 100191, China}
\affiliation[b]{Department of Physics, Liaoning University, Shenyang 110036
, China}

\emailAdd{ yijiegood@gmail.com}

\emailAdd{ still200@gmail.com}

\emailAdd{liukuiyong@lnu.edu.cn}

\emailAdd{nophy0@gmail.com}

\abstract{

The relativistic corrections to the productions of double $ J/\psi$ ,
double $ \Upsilon$, and $J/\psi+\Upsilon $ at the Tevatron and the LHC
were investigated within the frame of nonrelativistic QCD. The ratios of short distance
coefficients between relativistic correction and leading order result for color singlet
and color octet states in large $p_T$ limit are approximately $-1,-11/3$,
respectively, for pair production. And for $J/\psi+\Upsilon$ process the ratio is $-11/6$ for the CO channel.
The K factors of relativistic corrections
for color singlet and color octet double $J/\psi$~ production are $-0.23$ and $-0.84$ with adopting
$v^2=0.23$. The relativistic corrections significantly dilute the difference
between the shape of color-singlet differential cross sections and the color-octet's at leading order.
Also our results show that the single parton scattering model may be enough to explain the LHCb data
for the $J/\psi$ pair production.

}

\begin{document}
\maketitle
\flushbottom

\section{Introduction}

As the most achievable effective field theory to study production cross sections and decay rates for heavy quarkonium, the nonrelativistic QCD (NRQCD) \cite{Bodwin:1994jh} has
been widely accepted. In the NRQCD approach, a heavy quarkonium
can be described as a superposition of a short distance
coefficients $Q\bar{Q}_n(^{2s+1}L_J)$ mutiplied by a long-distance NRQCD matrix element $<O^H_n(^{2s+1}L_J)>$, here,
s, L, J denote the quantum numbers for
the spin, the orbital angular momentum, the total angular momentum of the $Q\bar{Q}$ intermediate state, respectively, and n stands for the color for the heavy-quark-antiquark pair state. The short distance
coefficients indicate the creation or annihilation of a heavy quark pair can be calculated perturbatively with the
expansions by the strong coupling constant $\alpha_s$. And the long-distance NRQCD matrix element is the probability of the intermediate heavy-quark-antiquark pair to evolve into an observed quarkonium nonperturbatively.
In general, the intermediate state can be in a conventional color-singlet or a color-octet state.
By introducing the color-octet mechanism, one can get the infrared-safe calculations for the decay rates of P-wave quarkonium states\cite{Huang:1996fa,Huang:1996sw,Huang:1996cs,Bodwin:1992ye,
Brambilla:2008zg,Brambilla:2006ph,Brambilla:2002nu,Brambilla:2001xy} and
D-wave\cite{He:2008xb,He:2009bf,Fan:2009cj} decay widths. Besides, one may also explain the Tevatron
data on the surplus production of $J/\psi$ and $\psi '$ at large
$p_T$ by fitting the data and extracting the color-octet long-distance NRQCD matrix elements.

Leaving aside the successes NRQCD have achieved, there are still puzzles in the comparisons between the NRQCD predictions
and the experimental data. The initial one is the transversely polarized quarkonium prediction
at the Tevatron failing in confronting with experimental measurements. This puzzle is called a "smoking gun" for the color-octet production mechanism. And recent ones are the large
discrepancies between the leading order calculations of inclusive\cite{Cho:1996cg,Yuan:1996ep,Liu:2003jj} and exclusive\cite{Braaten:2002fi,Liu:2002wq} charmonia productions and the experimental data  \cite{Abe:2002rb,Abe:2004ww,Aubert:2005tj} at B factories. Now the
most achievable efforts to resolve these puzzles seem to be higher order corrections. The next-to-leading order (NLO) QCD corrections\cite{Zhang:2006ay,Wang:2011qg,Zhang:2008gp,Zhang:2009ym,
Gong:2009ng,Gong:2009kp,Ma:2008gq,Dong:2011fb,
Bodwin:2013ys}, relativistic corrections\cite{He:2007te,Jia:2009np,He:2009uf,Fan:2012dy,Fan:2012vw}, and ${\mathcal O}(\alpha_s v^2)$ correction \cite{Dong:2012xx,Li:2013qp} largely compensate the discrepancies
between theoretical values and experimental measurements at B factory. The B factories
analysis also show that the the color-octet matrix elements may be overestimated previously\cite{Chao:2012iv,Zhang:2009ym,Kniehl:1997fv}. And the contributions of higher order corrections for charmonium prediction and polarization at hadronic colliders are still significant
\cite{Campbell:2007ws,Gong:2008hk,Gong:2010bk,Gong:2008sn,Gong:2012ah,Gang:2012js,Chao:2012iv,Ma:2010vd,
Butenschoen:2012px,Gong:2012ug}.  There are more information in the Ref.\cite{Brambilla:2010cs,Brambilla:2004wf}.

Most recently, the LHCb  measured the $J/\psi$ pair production at a center-of-mass energy of
$\sqrt{s}=7TeV$, with an integrated luminosity of 35.2 $pb^{-1}$ and give the differential cross sections as a function of the $J/\psi$ pair invariant mass\cite{Aaij:2011yc}.
These predictions are thought to be useful to study the color-octet
mechanism in quarkonium production by distinguishing the color-siglet contribution and color-octet contribution at different $p_T$ region\cite{Barger:1995vx,Qiao:2002rh,Qiao:2009kg,Li:2009ug,Ko:2010xy}.
However, theoretical predictions within the single parton scattering mechanism are small.
So double parton scattering mechanism was introduced in Ref.\cite{Berezhnoy:2011xy,Kom:2011bd,Baranov:2011ch}.
The predictions combining the double parton scattering and the single parton scattering lead a better agreement with data than that with the single parton scattering alone.

However, as mentioned above, high order calculations including QCD and relativistic
corrections to both CS and CO states are essential for heavy quarkonium production
in the single parton scattering process. A complete next leading order calculations to double quarkonium production would be helpful
to understand the multiple quarkonium production mechanism, COM mechanism, and the double
parton scattering effects, at hadron colliders.

Based on this motivation, in this paper, we studied the effects of relativistic corrections
to both the CS channel and CO channel of production of double $ J/\psi$ , production of double
$ \Upsilon$ at the LHC and the Tevatron. And we also discussed the relativistic corrections to
the $J/\psi+\Upsilon$ process for a comparison.

This paper is organized as follows. In section II the NRQCD factorization and the
standard methods to calculate the short distance coefficients of relativistic corrections
are presented. In section III, we would give the numerical results and make a discussion.
Finally there is a summary.

\section{The method of the calculations on relativistic corrections in NRQCD}

The differential cross section of two heavy quarkonium production can be discribed
by convoluting the cross parton level sections in the parton model as the following expression:
\begin{eqnarray}
d\sigma\big(p+p(\bar{p}){\rightarrow}H_1+H_2+X\big)
=\sum_{a,b}{\int}dx_1dx_2{f_{a/p}(x_1)}{f_{b/p(\bar{p})}(x_2)}
d\hat{\sigma}(a+b{\rightarrow}H_1+H_2),
\end{eqnarray}
where $f_{a(b)/p(\bar{p})}(x_i)$ is the parton distribution function(PDF), and $x_i$
is the parton momentum fraction denoted the fraction parton carried from proton or antiproton.

Within the NRQCD framework, the parton level cross sections can be divided into two parts:
short distance coefficients and long distance matrix elements:\cite{Bodwin:1994jh}
\begin{equation}
\label{eq:factorization} \hat{\sigma}(a+b{\rightarrow}H_1+H_2)=\sum_{mn}\frac{F_{mn}(ab)}{m_{Q1}^{d_m-4}m_{Q2}^{d_n-4}}\langle0|\mathcal{O}_m^{H_1}|0\rangle\langle0|\mathcal{O}_n^{H_2}|0\rangle.
\end{equation}
On the right-hand side of the equation, the cross section is expanded to sensible
fock states. $F_{mn}$ noted by the subscript $m, n$, i.e. short distance coefficients,
which can be obtained matching with the perturbative QCD,
describe the process that  produces
intermediate $Q\bar{Q}$ pairs in short range before harmonization to the physical meson state.
$\langle0|\mathcal{O}_{m,n}^{H}|0\rangle$, the long distance matrix elements, represent the
harmonization process that $Q\bar{Q}$
evolutes to the CS final state by emitting soft gluons. The long matrix elements can be extracted
from the experiment data\cite{Fan:2009zq,Guo:2011tz}, determined by potential model\cite{Bodwin:2007fz}
or lattice calculations\cite{Bodwin:1996tg,Bodwin:2001mk,Bodwin:2005gg}. $\mathcal{O}_{m,n}^{H}$ are
local four fermion operators. The factor of $m_Q^{d_{m,n-4}}$ is introduced to make
$F_{mn}$ dimensionless.

In this thesis, our calculations only refer to the S-wave spin-triplet quarkonium ($J/\psi$,$\Upsilon$)
and its wave function looks like
\begin{equation}
 |H\rangle=\mathcal{O}(1)|Q\bar{Q}(^3S_1^{[1]})\rangle + \mathcal{O}(v)|Q\bar{Q}(^3P_J^{[8]})g\rangle + \mathcal{O}(v^2)|Q\bar{Q}(^1S_0^{[8]})gg\rangle + \mathcal{O}(v^2)|Q\bar{Q}(^3S_1^{[8]})gg\rangle +\dots.
\end{equation}
So the fock state expansion of Eq.(\ref{eq:factorization}) may involve different combinations of
these states in above expressions.

For the computations of short-distance coefficients, there are different processes contributing
to the two meson production at parton level. In this paper, we just consider the gluon fusion
process which is the dominate one of the parton level processes, i.e.
$g(k_1)+g(k_2){\rightarrow}Q_1\bar{Q}_1+Q_2\bar{Q}_2{\rightarrow}H_1(p_1)+H_2(p_2)$.
Then  the Lorentz-invariant Mandelstam variables at parton level are writen by
(Here we omit the hats on the top of the $s$, $t$, $u$ for simplification)
\begin{eqnarray}\label{manvar}
s&=&(p_1+p_2)^2=(k_1+k_2)^2,\nonumber\\
t&=&(k_1-p_1)^2=(k_2-p_2)^2,\nonumber\\
u&=&(k_2-p_1)^2=(k_1-p_2)^2,
\end{eqnarray}
and they satisfy
\begin{equation}\label{stu}
s+t+u=M_{H_1}^2+M_{H_2}^2.
\end{equation}

In the LO computation, the momenta of the heavy quark/antiquark in the intermediate
$Q\bar{Q}$ pair are set to half of the meson momentum. In an orthodox method to deal
with the relativistic corrections, the amplitudes would be expanded by the rest relative
momentum of the $Q\bar{Q}$ around the quark/antiquark mass. In this paper, we also adhere
to this strategy. We first define the momentum of heavy quark/antiquark as $p_Q$/$p_ {\overline{Q}}$.
In a arbitrary frame, they can be written as the following expressions\cite{He:2007te,Braaten:1996jt,Ma:2012ex}:
\begin{eqnarray}
{p_Q}&=&\frac{1}{2}p+q,
\nonumber\\
{p_ {\overline{Q}}}&=&\frac{1}{2}p-q.
\end{eqnarray}
where $p$ boosts from the momentum of the quarkonium in rest frame of the meson, etc.
$(2E_q,\vec{0})$ and q boosts from $(0,\vec{q})$.  Here, $E_q=\sqrt{m_q^2+|\vec{q}|^2}$
is the rest energy of quark(antiquark).
$|\vec{q}|=m_Q v$ is a lorentz scalar and $v$ is the three relative velocity of quark or
the antiquark in the rest frame.
We can estimate $v$ from the Gremm-Kapustin relation in weak-coupling regime\cite{Gremm:1997dq,Bodwin:2003wh}
\begin{equation}\label{gkrelation}
v^2=v_1^2=v_8^2=\frac{M_{Q\overline{Q}}-2m_Q^{pole}}{2m_Q^{QCD}}.
\end{equation}
where subscript "1" and "8" represents the pair of quark and antiquark is in a CS or CO state.
And $m_Q^{QCD}$ is the mass of quark in NRQCD actions and $m_Q^{pole}$ is the pole mass of quark.
Taking $M_{Q\overline{Q}}$, $m_Q^{pole}$, and $m_Q^{QCD}$ as the input parameters, we can estimate
the value of $v$. For example, in this paper, we select $M_{J/\psi}=3.1GeV$ and $m_c^{QCD}=m_c^{pole}=1.39$ ,
and then get $v^2\sim0.23$.
And $M_{\Upsilon}=9.4GeV$ and $m_b^{QCD}=m_b^{pole}=4.57$ ,and then get $v^2\sim0.1$.
With the GK relation
\begin{equation}\label{matrixrelation}
v^2=\frac{\langle0|\mathcal{P}^{H(^{2s+1}L_J)}|0\rangle}
{m_Q^2\langle0|\mathcal{O}^{H(^{2s+1}L_J)}|0\rangle}[1+\mathcal{O}(v^4)].
\end{equation}
one may use the above values of $v$ in estimation of the sizes of next leading order LDMEs.

Taking advantage of the approximate relation $M_H=2E_q$ and Eqs.(\ref{manvar},\ref{stu}),
the Mandelstam variables may be expanded by the relative momentum $|\vec{q}|$. s is the beam energy
and is $|\vec{q}|$ independence. To expand $t,u$, we write they in the frame of gluon mass. Then the
expressions expanding to next order may be given as
\begin{eqnarray}\label{tuexpansion2}
t(|\vec{q}_{1}|,|\vec{q}_{2}|)&=&2E_{q1}^2+2E_{q2}^2-\frac{s}{2}-\frac{1}{2}\lambda^{\frac{1}{2}}\big(s,(2E_{q1})^2,(2E_{q2})^2\big)
\nonumber\\
&=&t(0,0)+Ft(2m_{Q1},2m_{Q2})|\vec{q}_{1}|^2+Ft(2m_{Q2},2m_{Q1})|\vec{q}_{2}|^2+\mathcal{O}(v^4),
\nonumber\\
u(|\vec{q}_{1}|,|\vec{q}_{2}|)&=&2E_{q1}^2+2E_{q2}^2-\frac{s}{2}-\frac{1}{2}\lambda^{\frac{1}{2}}\big(s,(2E_{q2})^2,(2E_{q2})^2\big)
\nonumber\\
&=&u(0,0)+Fu(2m_{Q1},2m_{Q2})|\vec{q}_{1}|^2+Fu(2m_{Q2},2m_{Q1})|\vec{q}_{2}|^2+\mathcal{O}(v^4).
\end{eqnarray}
where $\lambda(x,y,z)=x^2+y^2+z^2-2(x y+x z+ y z)$ and
\begin{eqnarray}
 Ft(x,y)&=&\frac{4\big\{[t(0,0)+u(0,0)](t(0,0)+y^2)-2y^2[t(0,0)+x^2]\big\}}{s[s-2(x^2+y^2)]+(x^2-y^2)^2]},\nonumber\\
 Fu(x,y)&=&\frac{4\big\{[t(0,0)+u(0,0)](u(0,0)+y^2)-2y^2[u(0,0)+x^2]\big\}}{s[s-2(x^2+y^2)]+(x^2-y^2)^2]}.
\end{eqnarray}
For the case that the final two mesons are identical particals, the expanding expressions would be simplified to
\begin{eqnarray}\label{tuexpansion1}
t(\vec{q}_{1},\vec{q}_{2})=t(0,0)-\frac{4\big[t(0,0)+4m_{Q}^2\big]}{s-16m_{Q}^2}(|\vec{q}_{1}|^2+|\vec{q}_{2}|^2)+\mathcal{O}(v^4),
\nonumber\\
u(\vec{q}_{1},\vec{q}_{2})=u(0,0)-\frac{4\big[u(0,0)+4m_{Q}^2\big]}{s-16m_{Q}^2}(|\vec{q}_{1}|^2+|\vec{q}_{2}|^2)+\mathcal{O}(v^4).
\end{eqnarray}

The short distance coefficients can be evaluated by matching the computations
of perturbative QCD and NRQCD:
\begin{eqnarray}
d\hat{\sigma}\Big|_{pert~QCD}
&&=\sum_{mn}\frac{F_{mn}}{m_{Q1}^{d_m-4}m_{Q2}^{d_n-4}}\langle0|\mathcal{O}_m^{Q1\bar{Q1}}|0
\rangle\langle0|\mathcal{O}_n^{Q2\bar{Q2}}|0
\rangle\Big|_{pert~NRQCD}.
\end{eqnarray}
The covariant projection operator method should be adopted to
compute the expression on the left-hand side of the equation.
Using this method, spin-singlet and spin-triplet combinations of
spinor bilinear in the amplitudes can be written in covariant form.
For instance, we give the spin-triplet projection,
\begin{eqnarray}
&\sum_{s\bar{s}}v(s)\bar{u}(\bar{s})\langle\frac{1}{2},s;\frac{1}{2},\bar{s}|1,S_z\rangle
=\frac{1}{\sqrt{2}(E_q+m_Q)}(\frac{\slashed{P}}{2}-\slashed{q}-m_Q)\frac{\slashed{P}-2E_q}{4E_q}
\slashed{\epsilon}(\frac{\slashed{P}}{2}+\slashed{q}+m_Q),
\end{eqnarray}
where $\epsilon$ denotes the polarization vector of the spin-triplet state.
Furthermore in our calculation the Dirac spinors are normalized as $\bar{u}u=-\bar{v}v=2m_Q$.
Then the differential cross section satisfies:
\begin{eqnarray}\label{perQCD}
&&d\hat{\sigma}(g+g{\rightarrow}Q_1\bar{Q}_1({}^{2s_1+1}L_{J1}^{[1,8]})+Q_2\bar{Q}_2({}^{2s_2+1}L_{J2}^{[1,8]}))\nonumber\\
&&{\sim}\overline{\sum}|\mathcal{M}(g+g{\rightarrow}
(Q_1\bar{Q}_1)(^{2s_1+1}L_{J1}^{[1,8]})+(Q_2\bar{Q}_2)(^{2s_2+1}L_{J2}^{[1,8]}))|^2
\nonumber\\
&&\times{\langle0|\mathcal{O}^{H_1}(^{2s_1+1}L_{J1})|0\rangle\langle0|\mathcal{O}^{H_2}(^{2s_2+1}L_{J2})|0\rangle}.
\end{eqnarray}
where the amplitude $\mathcal{M}$ is calculated by the following expression
\begin{eqnarray}
&&\mathcal{M}(g+g{\rightarrow}
(Q_1\bar{Q}_1)(^{2s+1}L_J^{[1,8]})+(Q_2\bar{Q}_2)(^{2s+1}L_J^{[1,8]}))
\nonumber\\
&&=\sum_{s_1\bar{s}_1s_2\bar{s}_2}\sum_{ijkl}\langle\frac{1}{2}, s_1;\frac{1}{2},
\bar{s}_1|J_1, J_{1z}\rangle\langle\frac{1}{2}, s_2;\frac{1}{2},
\bar{s}_2|J_2, J_{2z}\rangle\langle3i;\bar{3j}|1, 8a\rangle\langle3k;\bar{3l}|1, 8b\rangle \nonumber \\
&&\hspace{1cm}\mathcal{A}
(a+b{\rightarrow}Q_1^i+\bar{Q}_1^j+Q_2^i+\bar{Q}_2^j).\nonumber
\end{eqnarray}

 Next, we first expand the amplitude in powers of $|\vec{q}_1|$ and $|\vec{q}_2|$ to
 next order in $v^2$ (Here we only give the S-wave case and it is the only case using in our consequent calculations)
\begin{eqnarray}
&&\mathcal{M}(g+g\rightarrow(Q_1\bar{Q}_1)+(Q_2\bar{Q}_2))=
(\frac{m_{Q1}}{E_{q1}})^{1/2}(\frac{m_{Q2}}{E_{q2}})^{1/2}\mathcal{M}(q_1,q_2,E_{q1},E_{q2})\nonumber\\
&&=\mathcal{M}(0,0,m_{Q1},m_{Q2})+\big\{\frac{1}{2}q_1^{\alpha}q_1^{\beta}\frac{\partial^{2}
\mathcal{M}}{\partial q_1^{\alpha}\partial
q_1^{\beta}}+\frac{1}{2}q_2^{\alpha}q_2^{\beta}\frac{\partial^{2}
\mathcal{M}}{\partial q_2^{\alpha}\partial
q_2^{\beta}}
\nonumber\\
&&+\frac{|\vec{q}_1|^2}{2m_{Q1}}\frac{\partial\big[(m_{Q1}/E_{q1})^{1/2}\mathcal{M}\big]}{\partial E_{q1}}
+\frac{|\vec{q}_2|^2}{2m_{Q2}}\frac{\partial\big[(m_{Q2}/E_{q2})^{1/2}\mathcal{M}\big]}{\partial E_{q2}}\big\}\big(0,0,m_{Q1},m_{Q2}\big)
\nonumber\\
&&+ \mathcal{O}(|\vec{q}_1|^4) + \mathcal{O}(|\vec{q}_2|^4)\equiv\mathcal{M}^{(0)}+\mathcal{M}^{(2)}+\mathcal{O}(|\vec{q}_1|^4) + \mathcal{O}(|\vec{q}_2|^4).
\end{eqnarray}
The factor $\displaystyle(\frac{m_{Q}}{E_q})^{1/2}$ comes from the relativistic normalization
of the Fock state. For the S-wave case, $q^{\alpha}q^{\beta}=
\frac{1}{3}|\vec{q}|^{2}(-g^{\alpha\beta}+\frac{p_1^{\alpha}p_1^{\beta}}{p_1^{2}})
=\frac{1}{3}|\vec{q}|^{2}\Pi^{\alpha\beta}$ and the odd-power terms of $q$ vanish.
Subsequently, by multiplying the complex conjugate of the amplitude, the amplitude squared
up to the next order can be obtained:
\begin{eqnarray}
|\mathcal{M}|^{2}&=&\mathcal{M}^{0}\mathcal{M}^{0*}+(\mathcal{M}^{0}\mathcal{M}^{2*}+h.c.)+\mathcal{O}(v^4)
\nonumber\\
&=&\mathcal{M}^{0}\mathcal{M}^{0*}
\nonumber\\
&+&\frac{1}{6}|\vec{q}_1|^{2}\Pi^{\alpha\beta}
[\frac{\partial^{2}\mathcal{M}}{\partial q_1^{\alpha}\partial q_1^{\beta}}(0,0,m_{Q1},m_{Q2})\mathcal{M}^{0\ast}+h.c.]
+\frac{|\vec{q}_1|^2}{2m_{Q1}}\frac{\partial(m_{Q1}/E_{q1}\mathcal{M})}{\partial E_{q1}}(0,0,m_{Q1},m_{Q2})
\nonumber\\
&+&\frac{1}{6}|\vec{q}_2|^{2}\Pi^{\alpha\beta}
[\frac{\partial^{2}\mathcal{M}}{\partial q_2^{\alpha}\partial q_2^{\beta}}(0,0,m_{Q1},m_{Q2})\mathcal{M}^{0\ast}+h.c.]
+\frac{|\vec{q}_2|^2}{2m_{Q2}}\frac{\partial(m_{Q2}/E_{q2}\mathcal{M})}{\partial E_{q2}}(0,0,m_{Q1},m_{Q2})
\nonumber\\
&+&\mathcal{O}(v^4).
\end{eqnarray}
Form the expressions, one can see the relativistic corrections contain four parts to the
two quarkonum production. To evaluate them, Eqs. (\ref{tuexpansion2}~ \ref{tuexpansion1}) would be used.


\subsection{$J/\psi$($\Upsilon$) pair production}

In Ref.\cite{Qiao:2010kn,Ko:2010xy,Ko:2010vh}, the production of $J/\psi$ pair was
discussed in NRQCD. In the fock state expansion of the cross section, only two combinations
give the dominate contributions and other ones are suppressed in all the $p_T$ region,
i.e. $|^3S_1^{[1]}\rangle|^3S_1^{[1]}gg\rangle$ state and $|^3S_1^{[8]}g\rangle|^3S_1^{[8]}g\rangle$.
The former item gives dominate contribution at relative lower $p_T$ region and
the latter becomes important along with $ p_T$ increasing. In this paper,
our computations restrict relativistic corrections of these two combinations.
Therefore the differential cross section in Eq.(\ref{eq:factorization})
up to next order in $v$ takes the following form
\begin{eqnarray}\label{csuptov}
d\hat{\sigma}(g+g \rightarrow
J/\psi+J/\psi)&=&\frac{F({}^3S_1^{[1]}{}^3S_1^{[1]})}{m_c^4}{\langle0|\mathcal{O}^{J/\psi}({}^3S_1^{[1]})|0\rangle}
{\langle0|\mathcal{O}^{J/\psi}({}^3S_1^{[1]})|0\rangle}
\nonumber\\
&+&2\times\frac{G({}^3S_1^{[1]}{}^3S_1^{[1]})}{m_c^6}{\langle0|\mathcal{P}^{J/\psi}({}^3S_1^{[1]})|0\rangle}
{\langle0|\mathcal{O}^{J/\psi}({}^3S_1^{[1]})|0\rangle}
\nonumber\\
&+&\frac{F({}^3S_1^{[8]}{}^3S_1^{[8]})}{m_c^4}{\langle0|\mathcal{O}^{J/\psi}({}^3S_1^{[8]})|0\rangle}
{\langle0|\mathcal{O}^{J/\psi}({}^3S_1^{[8]})|0\rangle}
\nonumber\\
&+&2\times\frac{G({}^3S_1^{[8]}{}^3S_1^{[8]})}{m_c^6}{\langle0|\mathcal{P}^{J/\psi}({}^3S_1^{[8]})|0\rangle}
{\langle0|\mathcal{O}^{J/\psi}({}^3S_1^{[8]})|0\rangle}.
\end{eqnarray}
where $ F(_{\cdots})$, $G(_{\cdots})$ are the short distance coefficients and
for leading order and relativistic correction respectively. The factor 2 in the relativistic
terms is given because the contributions of relativistic correction for identified particle
are naturally equal. So we can calculate only the contribution of one $J/\psi$, then multiply
by 2 to get the total correction. The four-Fermion operators $\mathcal{O}^{H}(\Lambda)$
with the dimension-6 and the dimension-8 four-Fermion operators $\mathcal{P}^{H}(\Lambda)$
for the relativistic correction operators are defined as\cite{Bodwin:1994jh}
\begin{eqnarray}\label{operators}
\mathcal{O}^{H}({}^3S_1^{[1]})&=&\psi^{\dag} \vec{\sigma}   \chi
\cdot\chi^{\dag}\vec{\sigma}    \psi
\nonumber\\
\mathcal{O}^{H}({}^3S_1^{[8]})&=&\psi^{\dag} \vec{\sigma}T^a\chi
\cdot\chi^{\dag}\vec{\sigma} T^a\psi
\nonumber\\
\mathcal{P}^{H}({}^3S_1^{[1]})&=&\frac{1}{2}[\psi^{\dag}\vec{\sigma}
\chi\cdot\chi^{\dag}\vec{\sigma}(-\frac{i}{2}\overleftrightarrow{\mathbf{D}})^2\psi+h.c.]
\nonumber\\
\mathcal{P}^{H}({}^3S_1^{[8]})&=&
\frac{1}{2}[\psi^{\dag}\vec{\sigma}T^a
\chi\cdot\chi^{\dag}\vec{\sigma}T^a(-\frac{i}{2}\overleftrightarrow{\mathbf{D}})^2\psi+h.c.]
\end{eqnarray}
where $\chi$ and $\psi$ are the Pauli spinors describing anti quark
creation and quark annihilation respectively. $\sigma$ is the Pauli
matrices and $\mathbf{D}$ is the gauge-covariant derivative with
$\overleftrightarrow{\mathbf{D}}=\overrightarrow{\mathbf{D}}-
\overleftarrow{\mathbf{D}}$. $T$ is the $SU(3)$ color matrix.

The above analysis is also adequate to the production of $\Upsilon$ pair.
We use FeynArts\cite{Kublbeck:1990xc} to generate Feynman diagrams and amplitudes and
FeynCalc\cite{Mertig:1990an} handle the amplitudes. For double $J/\psi's $ and double
$\Upsilon's$ process there are 31 Feynman diagrams for color-singlet channel
and 72 Feynman diagrams for color-octet channel.
We will give the results of the short distance coefficients to leading order
and relativistic corrections in the Appendix.
Our leading order results to color singlet and  color octet channel are consistent with the ref \cite{Ko:2010xy}.

\subsection{$J/\psi+\Upsilon$ channel}

As the discussion in Refs.\cite{Ko:2010xy,Ko:2010vh}, the events to double $J/\psi$
under the luminosity of LHC is 70 and it is difficult to study COM by use of the process
of $J/\psi$ pair production for the LHC experiments as well as for the di-$\Upsilon$ pair
production. They study the process $pp\rightarrow{J/\psi+\Upsilon+X}$ and find it is nice probe to COM with enough events at LHC. To study relativistic corrections to
this process is a meaningful work.
At leading order of $O(\alpha_s)$, the CS channel $c\bar{c}(^3S_1^{[1]})+ b\bar{b}(^3S_1^{[1]})$
have no contributions. The CO channels include
$c\bar{c}(^3S_1^{[1]})+ b\bar{b}(^3S_1^{[8]})$, $c\bar{c}(^3S_1^{[8]})+
b\bar{b}(^3S_1^{[1]})$ and $c\bar{c}(^3S_1^{[8]})+ b\bar{b}(^3S_1^{[8]})$ combinations.
In our calculations, we neglect the relativistic corrections to
$\Upsilon$ production which is negligible compared with the corrections to
$J/\psi$. Therefore we have the obvious expression to next order in $v$
\begin{eqnarray}\label{csuptov}
d\hat{\sigma}(g+g \rightarrow
J/\psi+\Upsilon)&=&\frac{F({}^3S_1^{[1]}{}^3S_1^{[8]})}{m_c^2m_b^2}{\langle0|
\mathcal{O}^{J/\psi}({}^3S_1^{[1]})|0\rangle}
{\langle0|\mathcal{O}^{\Upsilon}({}^3S_1^{[8]})|0\rangle}
\nonumber\\
&+&\frac{G({}^3S_1^{[1]}{}^3S_1^{[8]})}{m_c^4m_b^2}{\langle0|\mathcal
{P}^{J/\psi}({}^3S_1^{[1]})|0\rangle}
{\langle0|\mathcal{O}^{\Upsilon}({}^3S_1^{[8]})|0\rangle}
\nonumber\\
&+&\frac{F({}^3S_1^{[8]}{}^3S_1^{[1]})}{m_c^2m_b^2}{\langle0|\mathcal
{O}^{J/\psi}({}^3S_1^{[8]})|0\rangle}
{\langle0|\mathcal{O}^{\Upsilon}({}^3S_1^{[1]})|0\rangle}
\nonumber\\
&+&\frac{G({}^3S_1^{[8]}{}^3S_1^{[1]})}{m_c^4m_b^2}{\langle0|\mathcal
{P}^{J/\psi}({}^3S_1^{[8]})|0\rangle}
{\langle0|\mathcal{O}^{\Upsilon}({}^3S_1^{[1]})|0\rangle}
\nonumber\\
&+&\frac{F({}^3S_1^{[8]}{}^3S_1^{[8]})}{m_c^2m_b^2}{\langle0|\mathcal
{O}^{J/\psi}({}^3S_1^{[8]})|0\rangle}
{\langle0|\mathcal{O}^{\Upsilon}({}^3S_1^{[8]})|0\rangle}
\nonumber\\
&+&\frac{G({}^3S_1^{[8]}{}^3S_1^{[8]})}{m_c^4m_b^2}{\langle0|\mathcal{P}^{J/\psi}({}^3S_1^{[8]})|0\rangle}
{\langle0|\mathcal{O}^{\Upsilon}({}^3S_1^{[8]})|0\rangle}.
\end{eqnarray}
The operators are defined in Eq.(\ref{operators}) and the results of the short
distance coefficients can also be found in the Appendix.

\section{The Numerical Result and Discussion}

Using the method of phase space integration in Ref. \cite{Fan:2009zq},
we can carry out the numerical calculation the cross section at the
LHC and Tevatron respectively. Here we ignore the relativistic correction come
from the phase space. Fortran program was used to calculate the integration.
We choose input parameters: the mass of charm quark $m_c=1.5GeV$ and bottom
quark $m_b=4.7GeV$; the factorization scale $\mu$ taken as transverse mass
$\mu=m_T=(4 {m_Q}^2+{p_T}^2)$ ; specifically we use the parton distribution
function CTEQ6L1\cite{Pumplin:2002vw}. The running coupling const $\alpha_s$
is evaluated by the formula of CTEQ6L1. We employ the following values for the
NRQCD matrix elements for $J/\psi$ and $\Upsilon$ \cite{Bodwin:2007fz,Braaten:1999qk,Kang:2007uv,Kramer:2001hh}:
\begin{eqnarray}
\langle0|\mathcal{O}^{J/\psi}({}^3S_1^{[1]})|0\rangle&=&1.3  GeV^3,
\nonumber\\
\langle0|\mathcal{O}^{J/\psi}({}^3S_1^{[8]})|0\rangle&=&3.9{\times}10^{-3}  GeV^3.
\end{eqnarray}
\begin{eqnarray}
\langle0|\mathcal{O}^{\Upsilon}({}^3S_1^{[1]})|0\rangle&=&9.2 GeV^3,
\nonumber\\
\langle0|\mathcal{O}^{\Upsilon}({}^3S_1^{[8]})|0\rangle&=&1.5{\times}10^{-1}  GeV^3.
\end{eqnarray}
As discussion in Sec.2, we use Eq.\eqref{matrixrelation} to estimate the relativistic LDMES.

\subsection{$J/\psi$($\Upsilon$) pair production}

Under the large $p_T$ limit approximation,
\begin{eqnarray}
-\frac{m^2}{u}\sim- \frac{m^2}{t}<\frac{m^2}{p_T^2}\sim0
\end{eqnarray}
where $m$ is the mass of $J/\psi$ or $\Upsilon$, we can expand the short distance
coefficients with $m$. The ratios of the short distance coefficient between LO (F)
and its relativistic correction (G) in lowest order in the expansion are:
\begin{eqnarray}
R({}^3S_{1}^{[1]}{}^3S_{1}^{[1]})\Big|_{p_T\gg m}=\frac{G({}^3S_1^{[1]}{}^3S_{1}^{[1]})}{F({}^3S_1^{[1]}{}^3S_{1}^{[1]})}\Big|_{p_T\gg m}&\sim&-1\nonumber \\
R({}^3S_{1}^{[8]}{}^3S_{1}^{[8]})\Big|_{p_T\gg m}=\frac{G({}^3S_1^{[8]}{}^3S_{1}^{[8]})}{F({}^3S_1^{[8]}{}^3S_{1}^{[8]})}\Big|_{p_T\gg m}&\sim&-\frac{11}{3}
\end{eqnarray}

 The ratio of ${}^3S_{1}^{[8]}$ is approximately $-11/3$ this ratio of color octet state
 is consistent with Ref.\cite{Bodwin:2003wh,Xu:2012am}. In large $p_T$ limit,
 the dominate contribution of this subprocess is $g^*\to c\bar c ({}^3S_{1}^{[8]})$.
 The propagator of virtual gluon $g^*$ is proportional to $1/E_q^2$.
 This term offers a factor of $-2$ to the ratio $R({}^3S_{1}^{[8]})$.

\begin{figure}[h]
\centering
\includegraphics[width=0.45\textwidth]{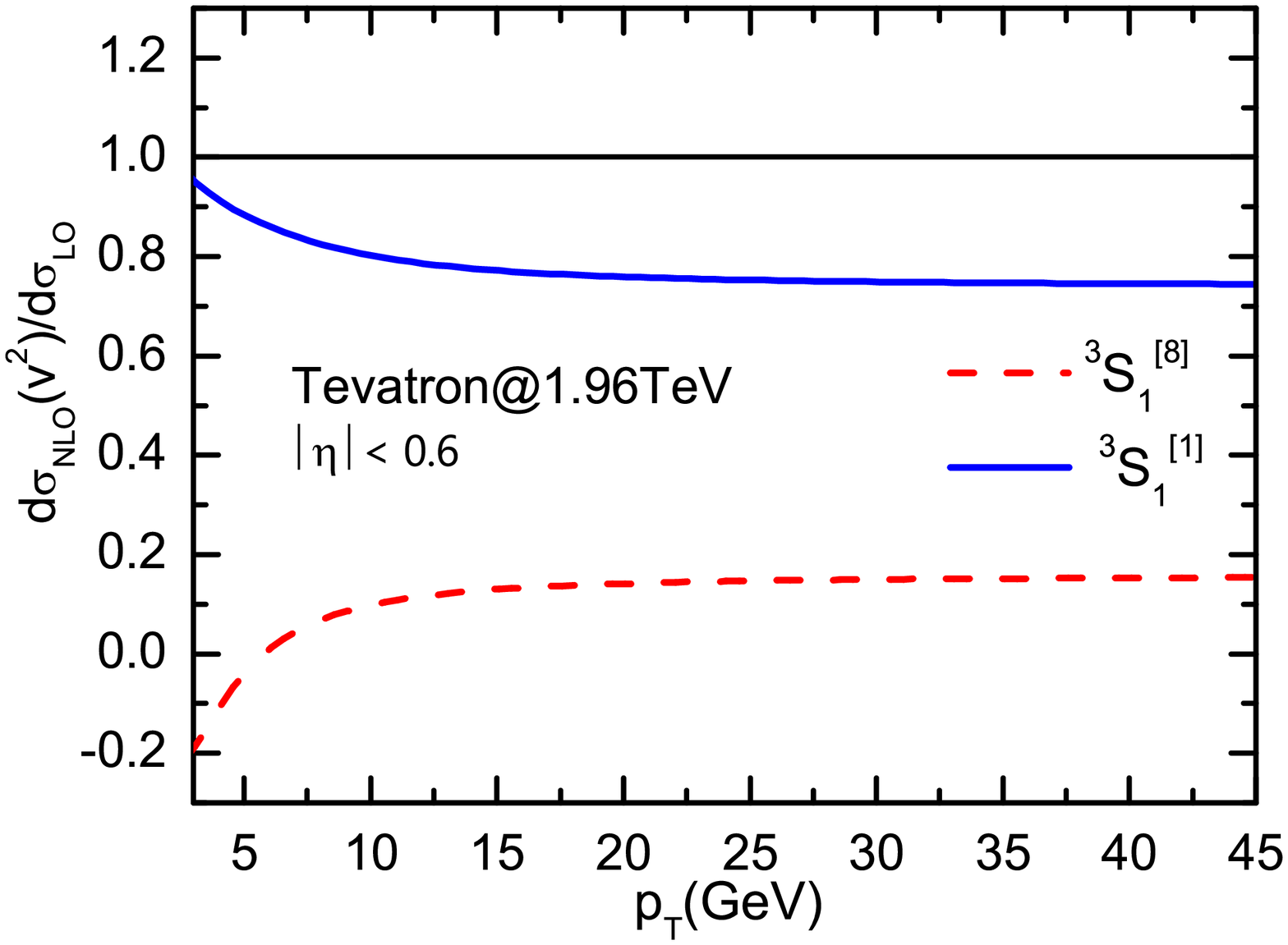}
\includegraphics[width=0.45\textwidth]{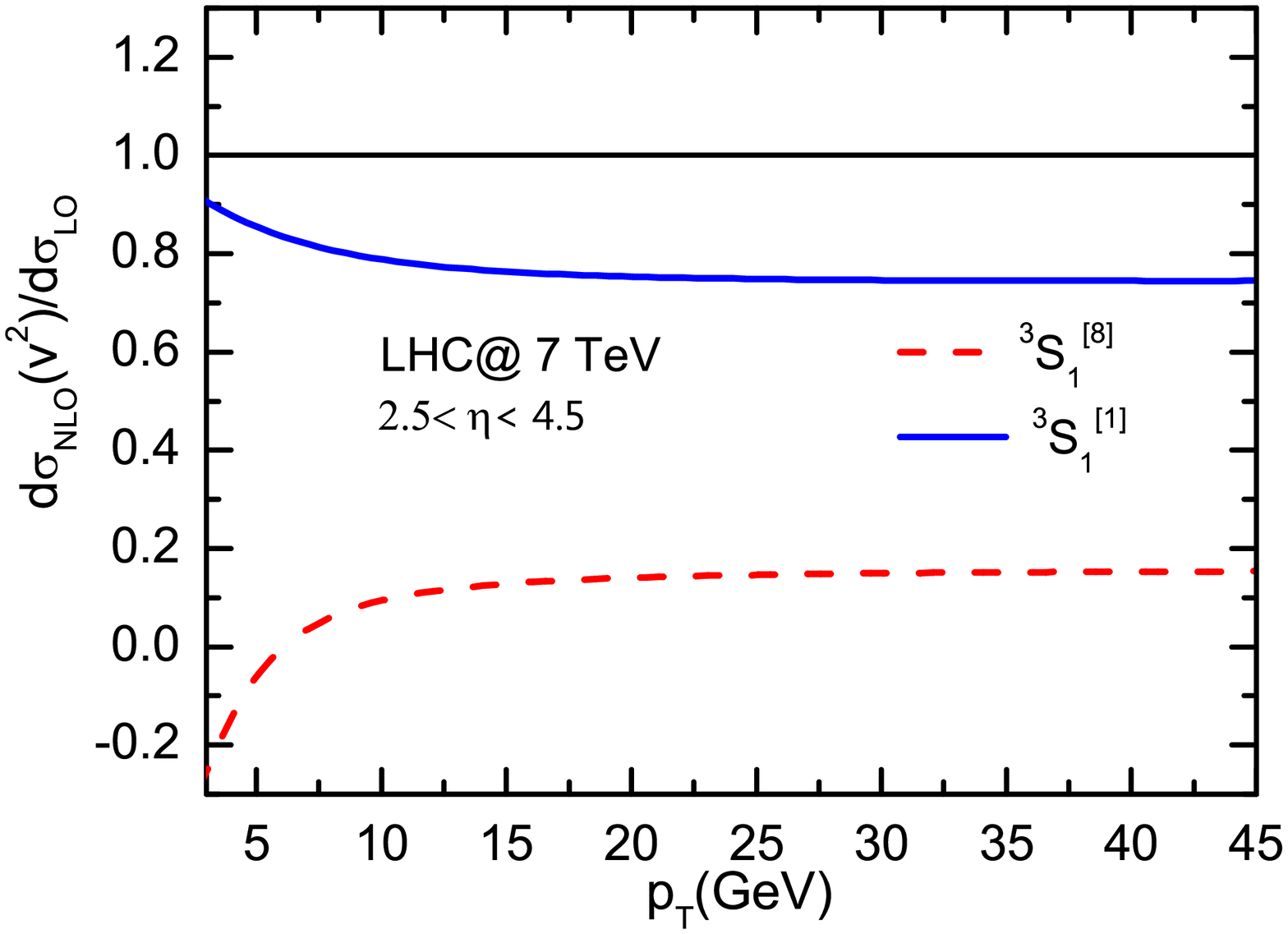}
\caption{\label{fig:rate}
The ratios of the short distance coefficient between LO $F$ and its relativistic correction G of ${p+p}\rightarrow{J/\psi+J/\psi}$ process at the Tevatron with $\sqrt{s}=1.96TeV$ and at the LHC with $\sqrt{s}=7TeV$ of $\sqrt{s}=14TeV$.
}
\end{figure}

\begin{figure}[h]
\centering
\includegraphics[width=0.49\textwidth]{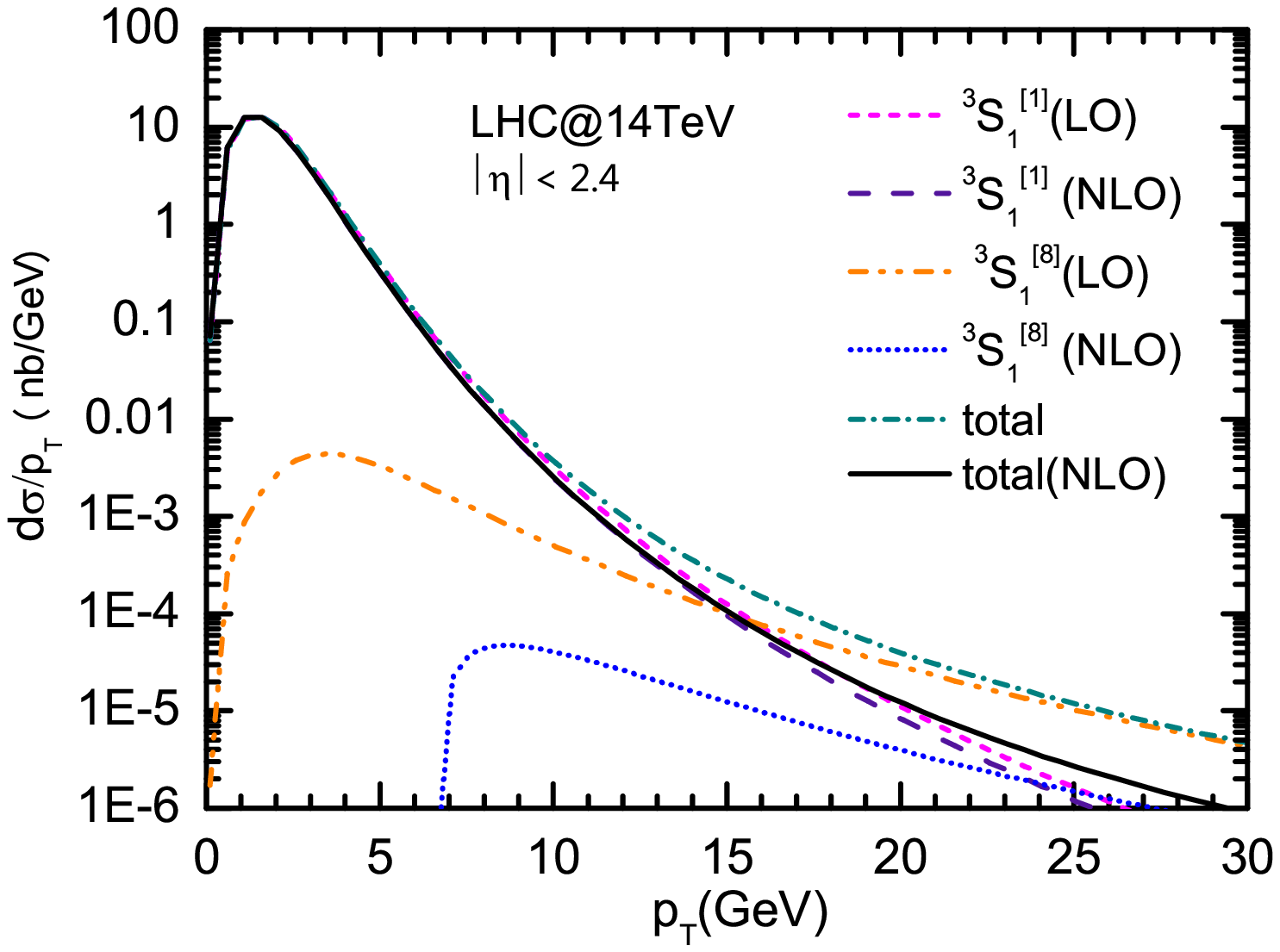}
\includegraphics[width=0.49\textwidth]{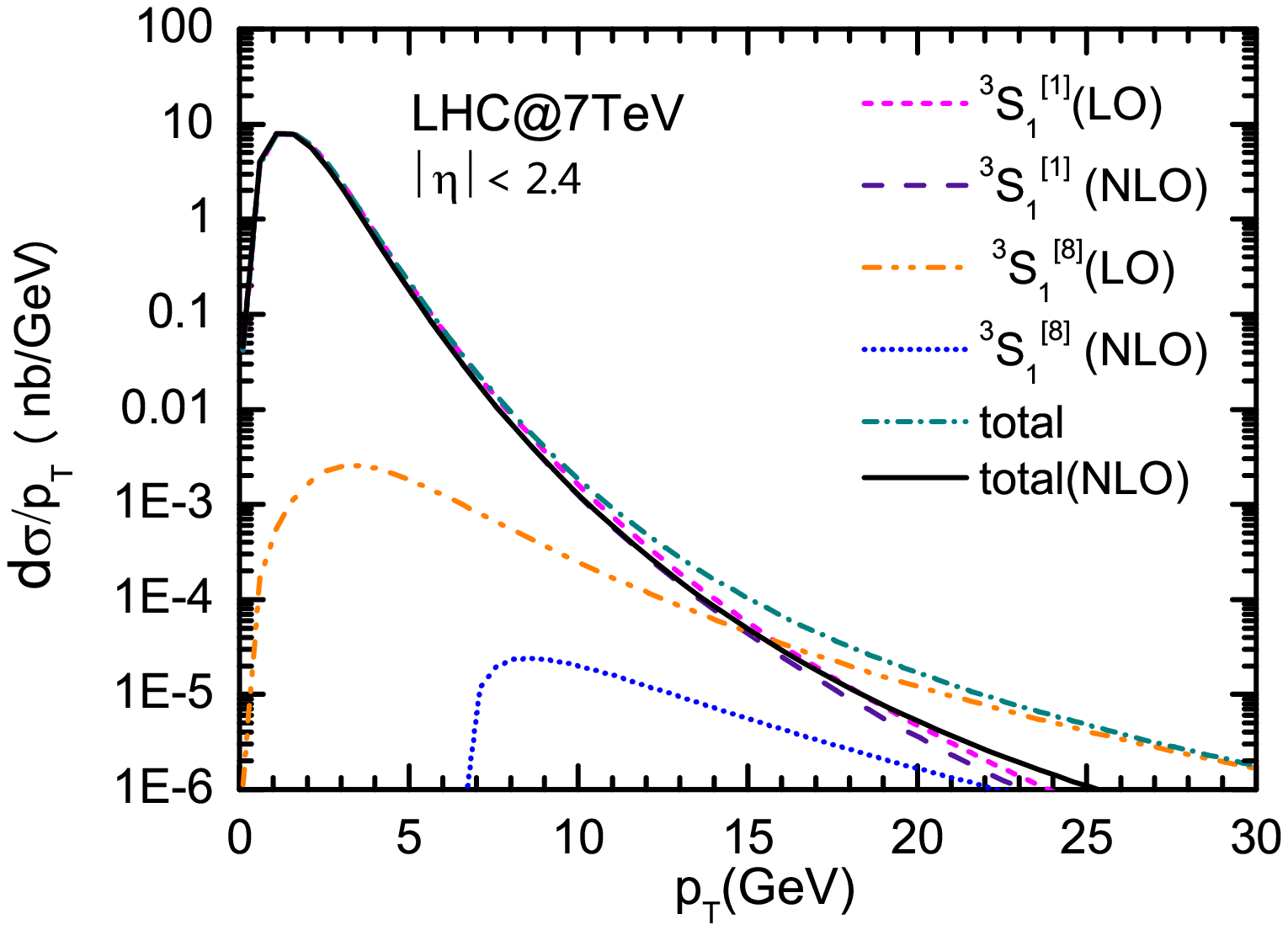}
\includegraphics[width=0.49\textwidth]{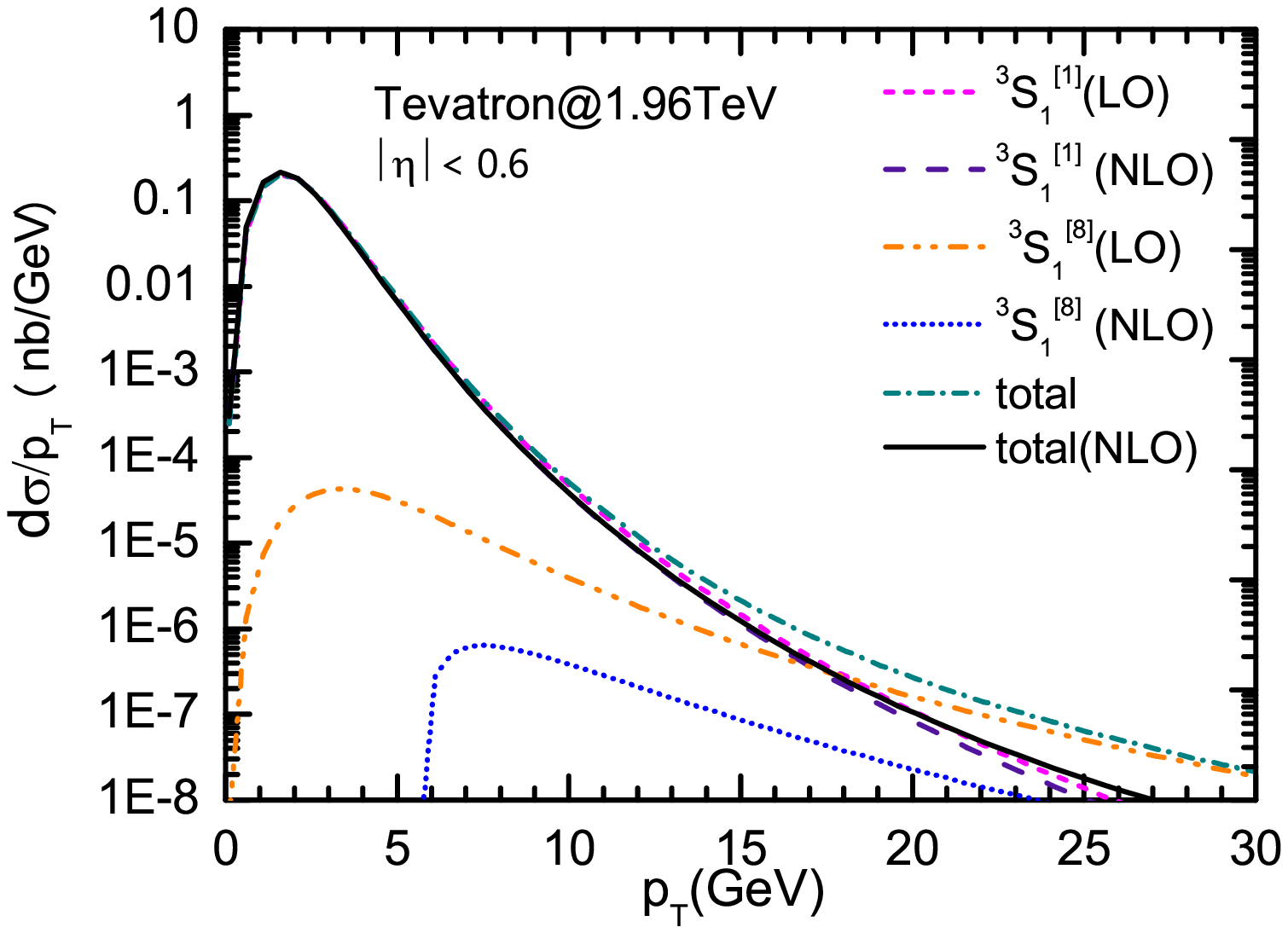}
\includegraphics[width=0.49\textwidth]{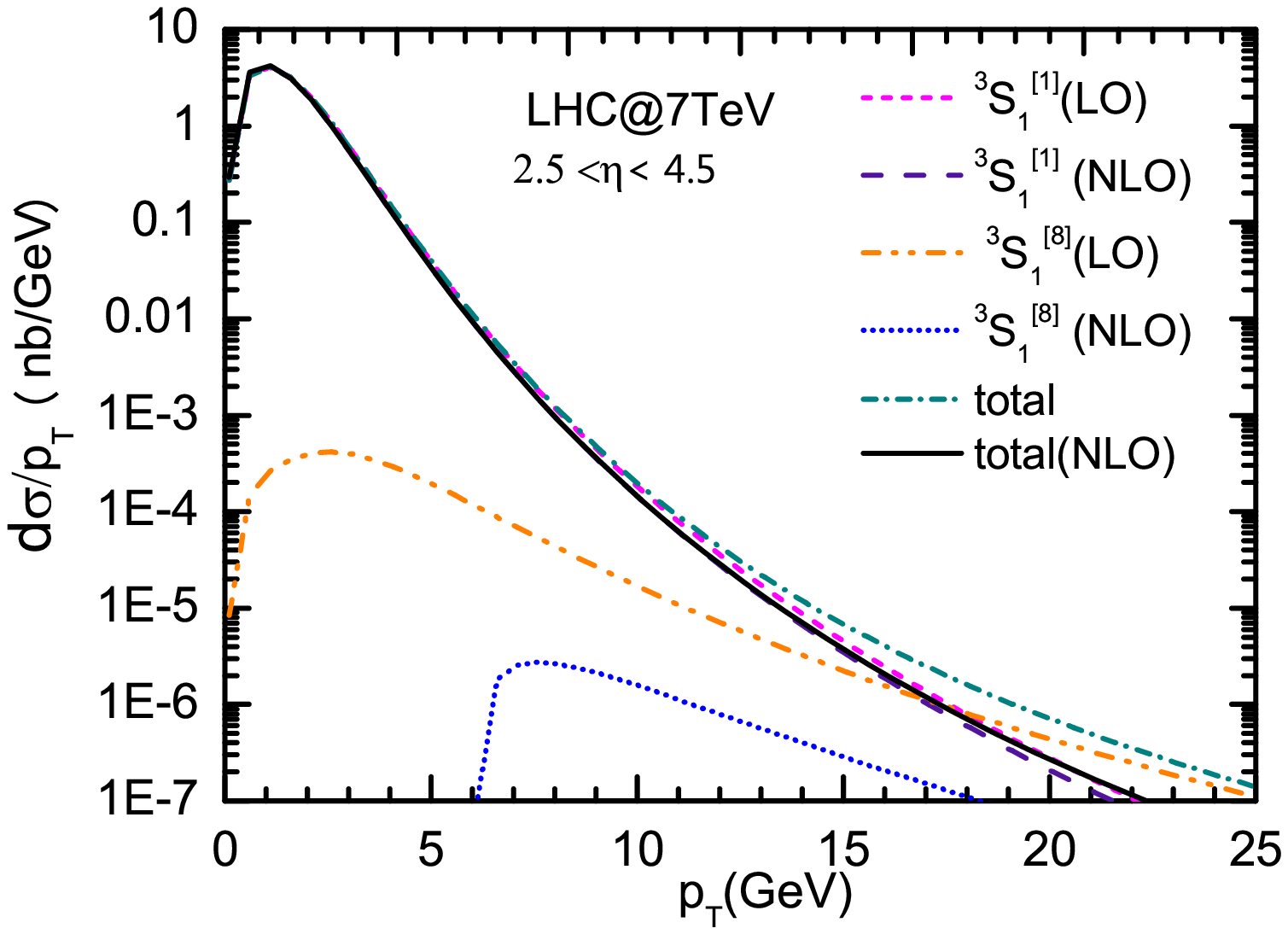}
\caption{\label{fig:cross}The ${p+p}\rightarrow{J/\psi+J/\psi}$ process  $p_t$ distribution of deferential cross section at the LHC and  Tevatron respectively.}
\end{figure}

\begin{figure}[h]
\centering
\includegraphics[width=0.49\textwidth]{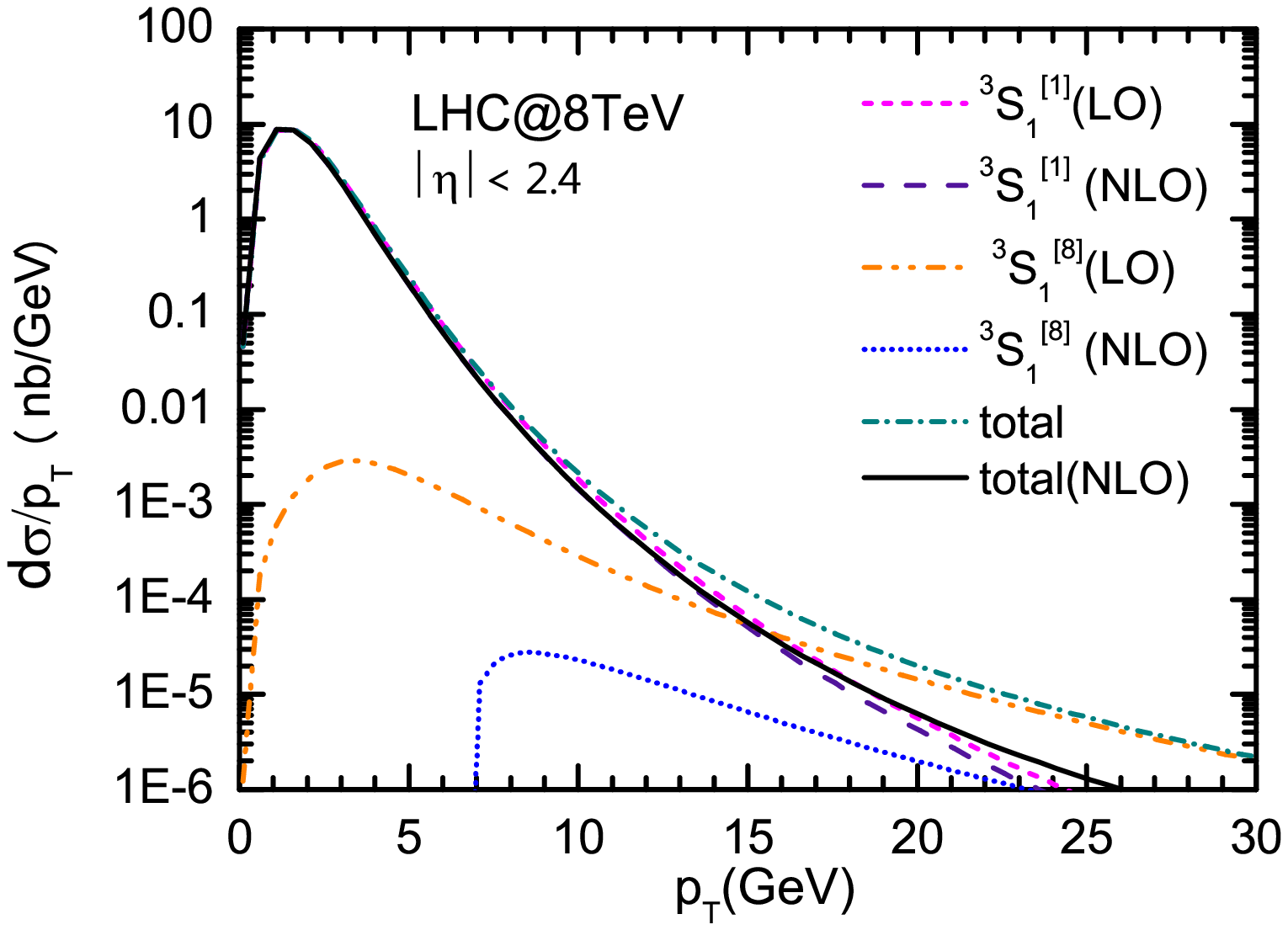}
\includegraphics[width=0.49\textwidth]{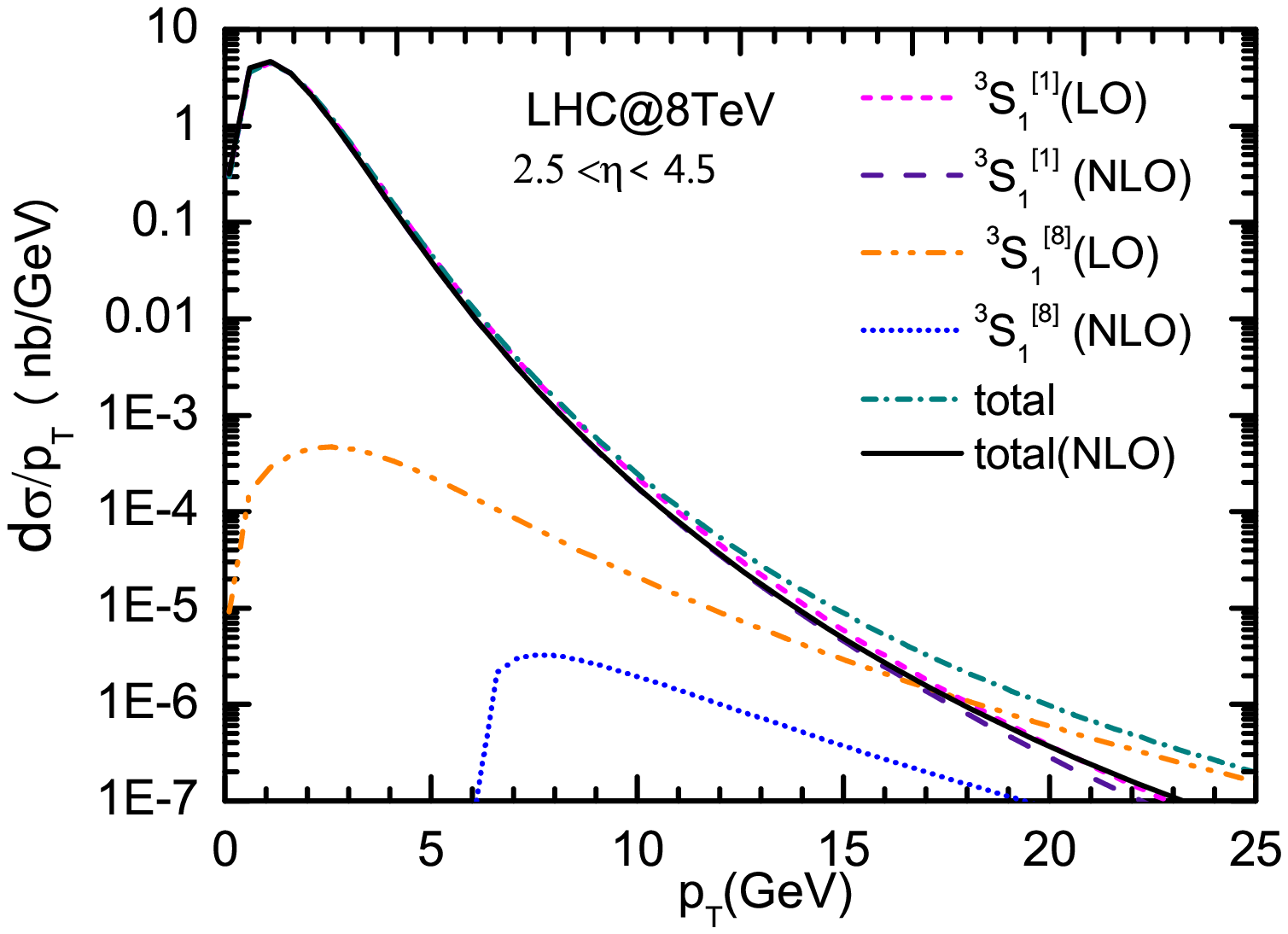}
\caption{\label{fig:cross2}The predicted ${p+p}\rightarrow{J/\psi+J/\psi}$ process  $p_t$ distribution of
deferential cross section at the LHC with  $\sqrt{s}=8TeV$ of $|y|<2.4$ and $2.5<y<4.5$ , respectively.}
\end{figure}

In Fig.\ref{fig:rate}, we present the ratios at the Tevatron with $\sqrt{s}=1.96TeV$
and at the LHC with $\sqrt{s}=7TeV$ or $\sqrt{s}=14TeV$ of $J/\psi$ and the curves
are very close at large $p_T$ and consistent with above analysis. In small $p_T$ region,
for CS, corrections are less than the values in large $p_T$ limit. On contrary,
there is a sharp suppress in small $p_T$ region for CO. That can obviously be shown
in Fig.\ref{fig:cross}, etc. differential cross sections of $p_T$.
The LO cross sections to CS and CO state suppressed by factor of $23\%$
and $84\%$ when involving the effects of relativistic corrections.
For the LO case, the CS and CO curves cross at about $p_T=14GeV$.
As $p_T$ increasing the contributions to CO state would surpass that
to CS. And this provide a probe to check the COM experimentally.
But in our calculations as shown in the diagrams, relativistic corrections
reduce the CO contributions largely and make the difference not apparently
between the CSM results and the COM results.
Furthermore,
we make a prediction of the cross sections at the LHCb with $\sqrt{s}=8TeV$
as shown in Fig.\ref{fig:cross2}, COM with relativistic corrections
contribute little to total cross sections that is similar with the cases mentioned above.
For total cross sections, COM effects are also not obviously, and this can be seen in Tab.\ref{total cross}.


\begin{figure}[h]
\centering
\includegraphics[width=0.45\textwidth]{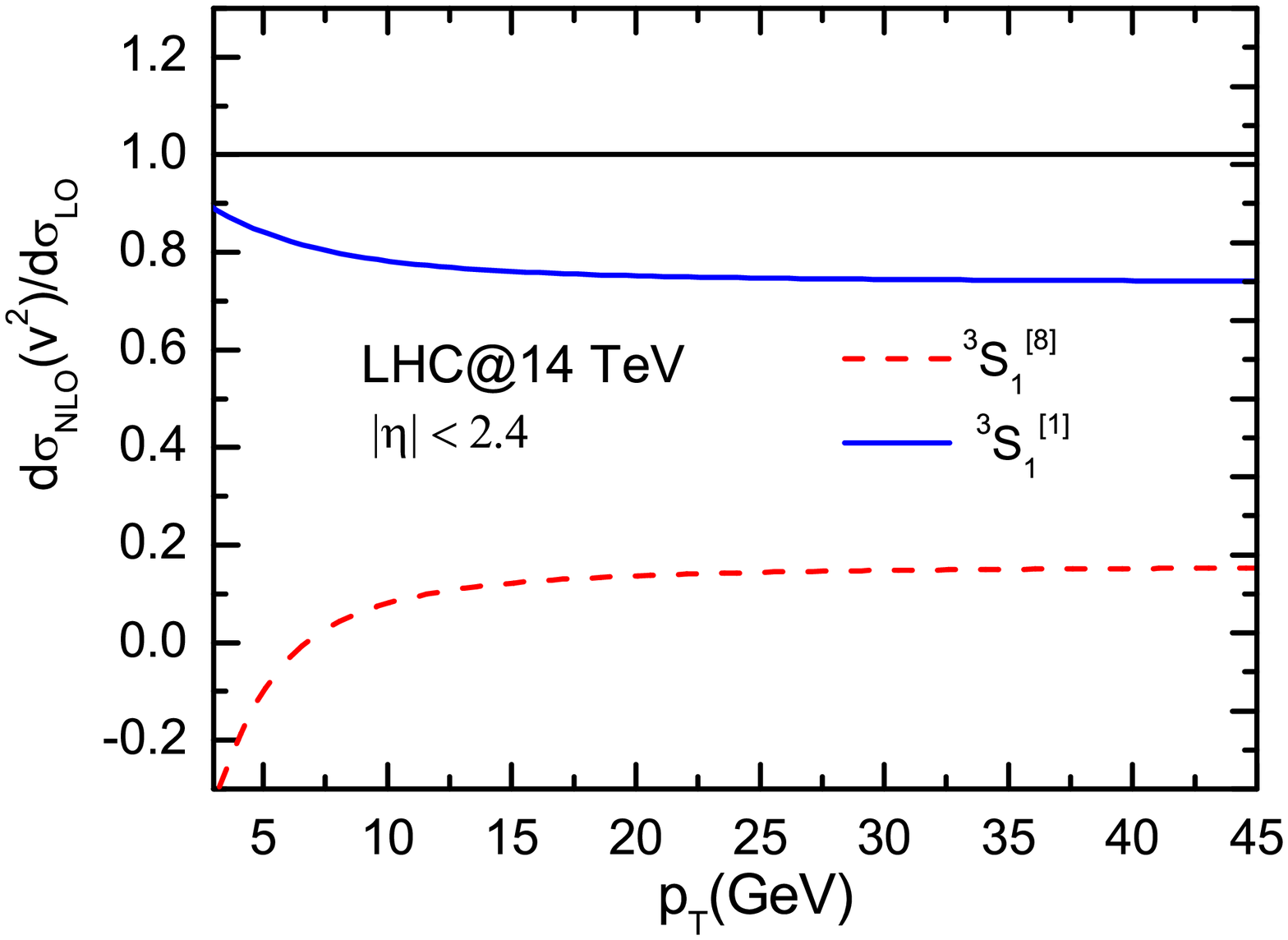}
\includegraphics[width=0.45\textwidth]{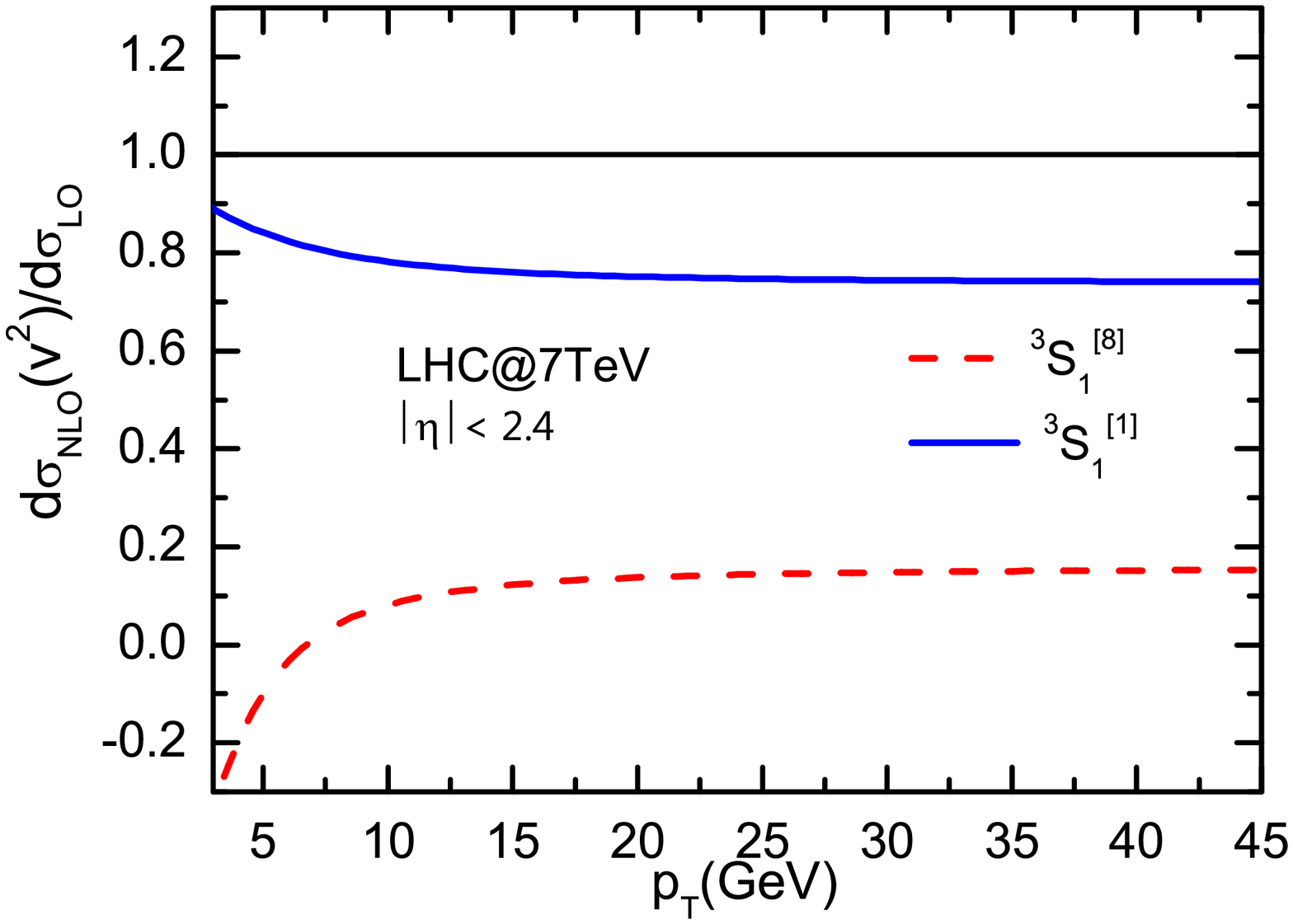}
\end{figure}

{
\tabcolsep=0.4mm
\begin{table}
 \begin{center}
\begin{tabular}{c|c|c|c|c|cp{2cm}|}
\hline\hline
$m_{c} $& $^3{S_1}^{[1]}$ LO & $^3{S_1}^{[1]}$ RC & $^3{S_1}^{[8]}$ LO& $^3{S_1}^{[8]}$ RC & total cross section  \\
(GeV)& (nb) & (pb) & (pb)&  (pb)& (nb) \\
\hline
1.60&5.18&$-13.90$&$1.63$&$-1.94$&5.16\\
\hline
1.55&6.92&$-22.12$&$2.20$&$-2.61$&6.89\\
\hline
1.50&9.31&$-34.68$&$3.00$&$-3.55$&9.28\\
\hline
1.45&12.65&$-53.09$&$4.13$&$-4.88$&12.60\\
\hline
1.40&17.34&$-80.96$&$5.73$&$-6.77$&17.26\\
\hline
\end{tabular}
\caption{\label{total cross} The total cross sections for color-singlet channel,
color-octet channel and their relativistic corrections at LHC with $\sqrt s=7TeV$,
integrating over $2<y<4.5$ and $p_T<10GeV$ and we choose charm quark mass for input
parameter from $1.4GeV$ to $1.6GeV$. "RC" at the head of the table denotes the relativistic contributions. }
\end{center}
 \end{table}}
In comparison of process of single $J/\psi$ production in which QCD corrections
are essential than relativistic corrections to CO states and give large contributions
to total cross sections. It is expected for QCD corrections to CS and CO states to give a large contributions.

\begin{figure}[h]
\centering
\includegraphics[width=0.49\textwidth]{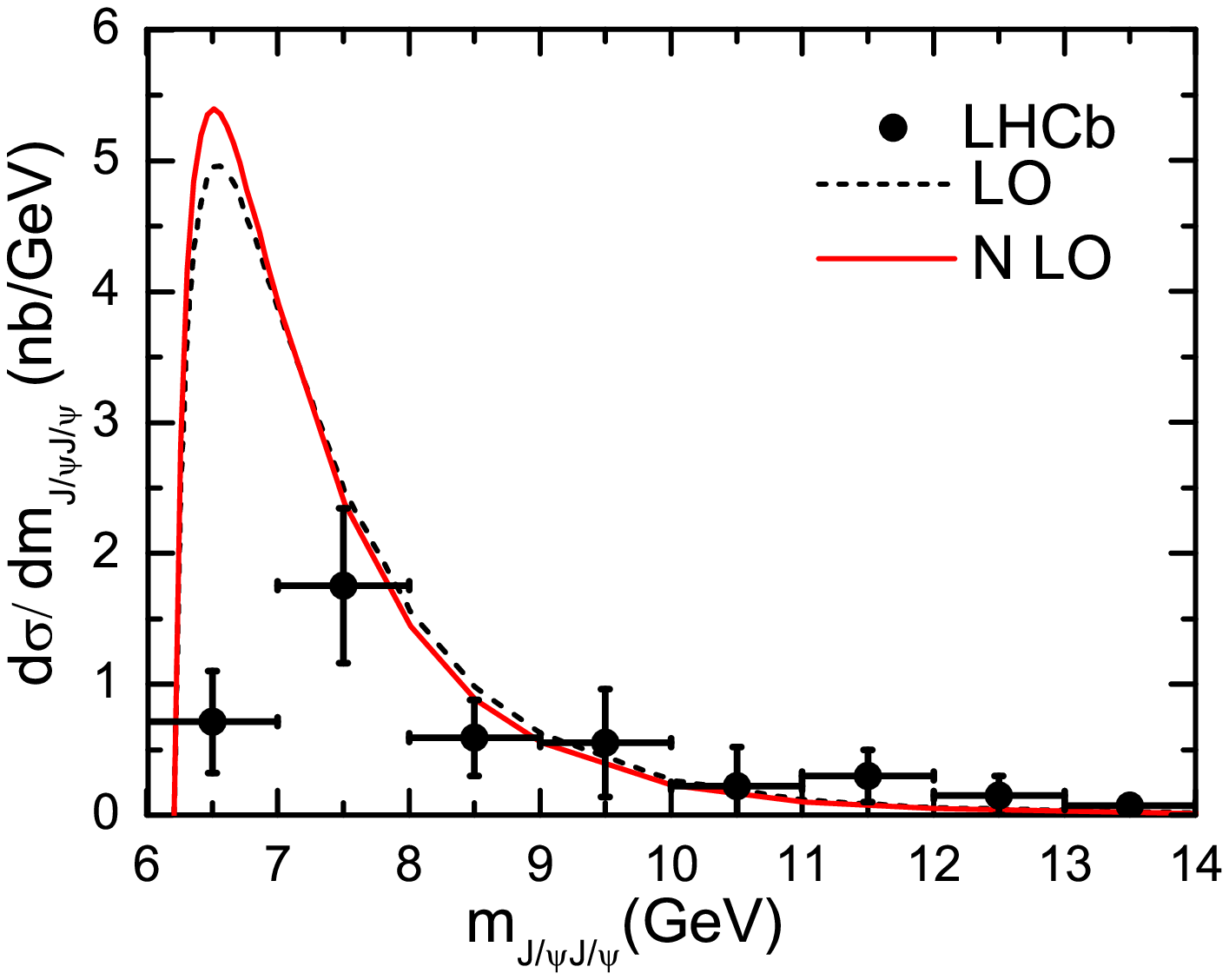}
\includegraphics[width=0.49\textwidth]{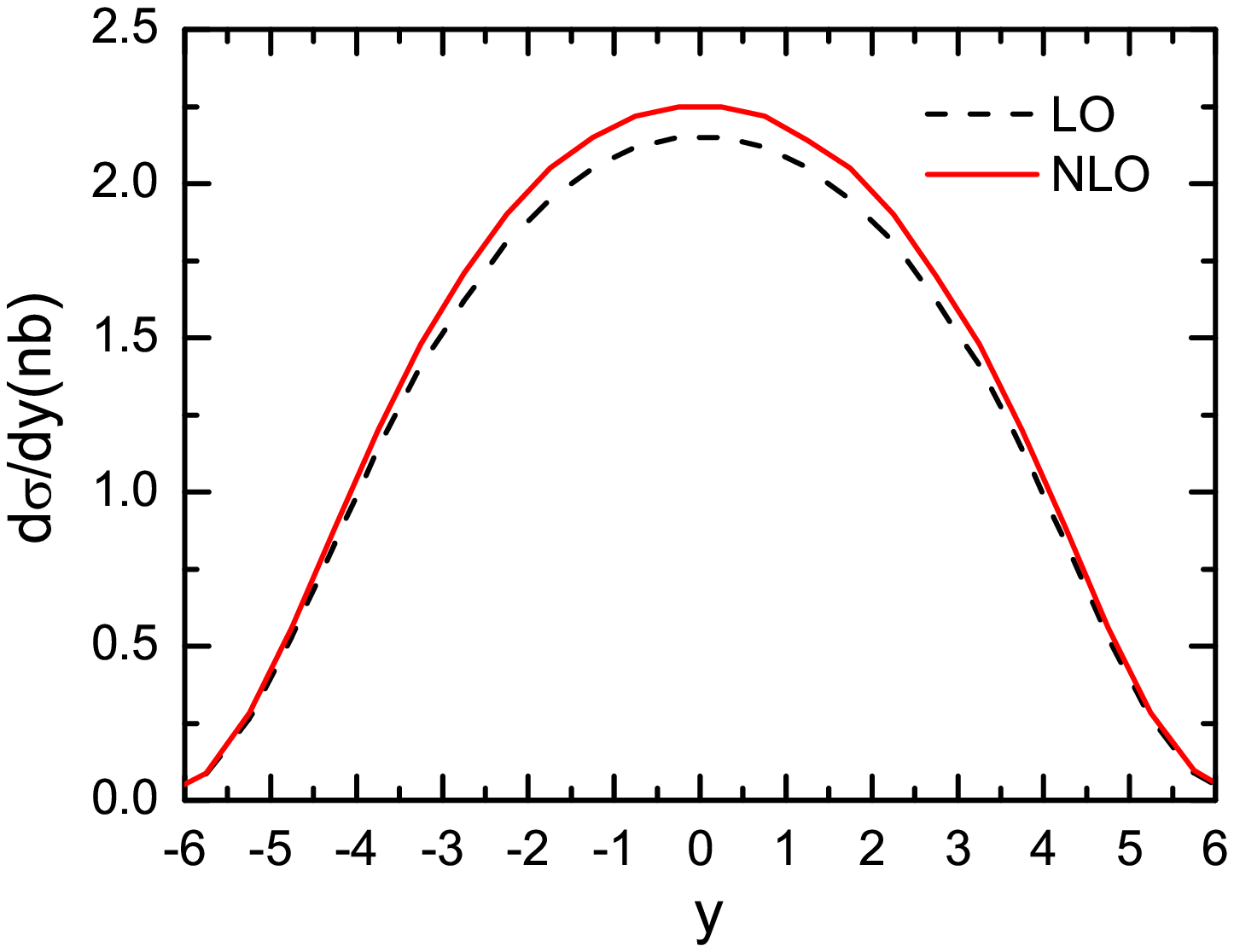}
\caption{\label{fig:data} Left figure is the differential cross section to double $J/\psi$ production as a function of the invariant mass $m_{J/\psi J/\psi}$ at LHC with $\sqrt s=7TeV$ and $2<y<4.5$, and dash and solid lines correspond to LO and NLO($v^2$) results, respectively. Right figure is for rapidity distribution, dash and solid lines also correspond to LO and NLO($v^2$), respectively. In both two figures, it were set to $M_{J/\psi}=3.1$.}
\end{figure}

To investigate the effects of relativistic corrections comparing with experiment,
we give the differential cross sections as functions of the invariant mass of double $J/\psi$
and of the rapidity in Fig.\ref{fig:data} to pair production with the same constraints
as the LHCb, i.e. $2<y<4.5$ and $\sqrt s=7TeV$. From the figures, one can see the main part of
the theoretical calculations concentrates in the low region of the invariant mass
with a sharp peak at about $6.5GeV$, but LHCb data shows a peak at $7\sim8GeV$.
And theoretical value of the peak is nearby $5.5nb/GeV$ which is two times large than the experimental peak value.
In the low region less than about $7.5GeV$, relativistic effects show a
positive correction to LO curve with a maximal enhanced factor 1.08.
In the region large than $7.5GeV$, relativistic effects suppress the LO curve but not obviously.
Here we only consider the contributions to direct pair production and the feed-down process
from excited state will contribute nearly $30\%$ to direct production rate\cite{Berezhnoy:2011xy}.
Taking the feed-down corrections into account, the theoretical prediction to differential
cross section within the invariant mass $6GeV\sim8.5GeV$ would be large than
the experimental values. This situation is very different from the calculations
in Refs.\cite{Berezhnoy:2011xy,Kom:2011bd}. They showed that the single parton
scattering contributions were too small to fit the experimental data and introduced
double parton scattering mechanism. In our calculations, only LO contributions
would be large enough comparing with the experimental data. But as they discussed,
single parton scattering result peaks at too low a invariant mass value and with
a too large value, and the relativistic corrections do not relieve the puzzle but
enhances the peak value. Recently, an soft-collinear effective theory(SCET)
has been developed and seem to provide a possibility to suppress the end point
peak considerably and move it from the end point\cite{Bauer:2000yr}.
Besides, may NLO QCD redia corrections give a large negative contribution
especially for the peak? It is expected the problem would be solved when involving these corrections.

\begin{figure}[h]
\centering
\includegraphics[width=0.49\textwidth]{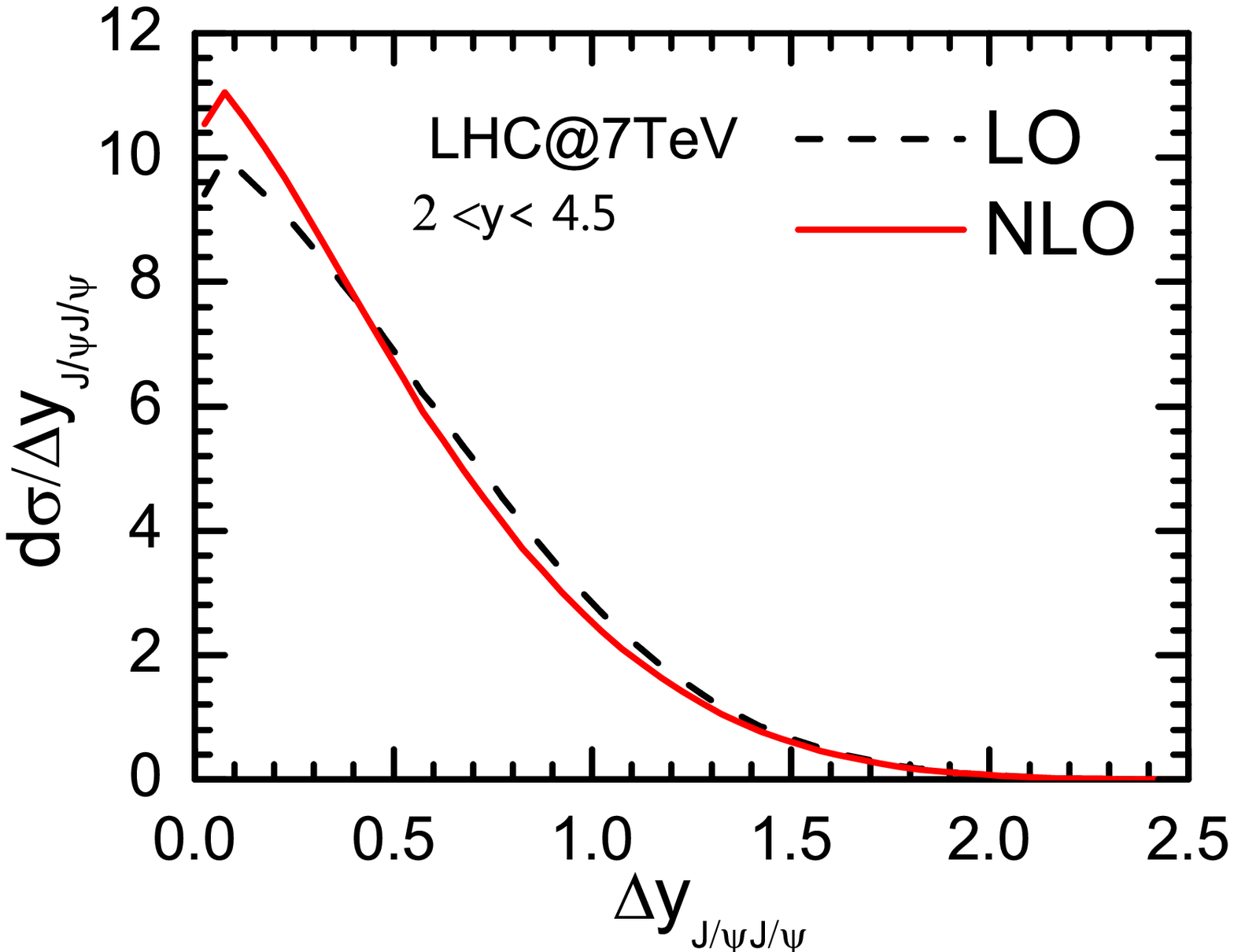}
\includegraphics[width=0.49\textwidth]{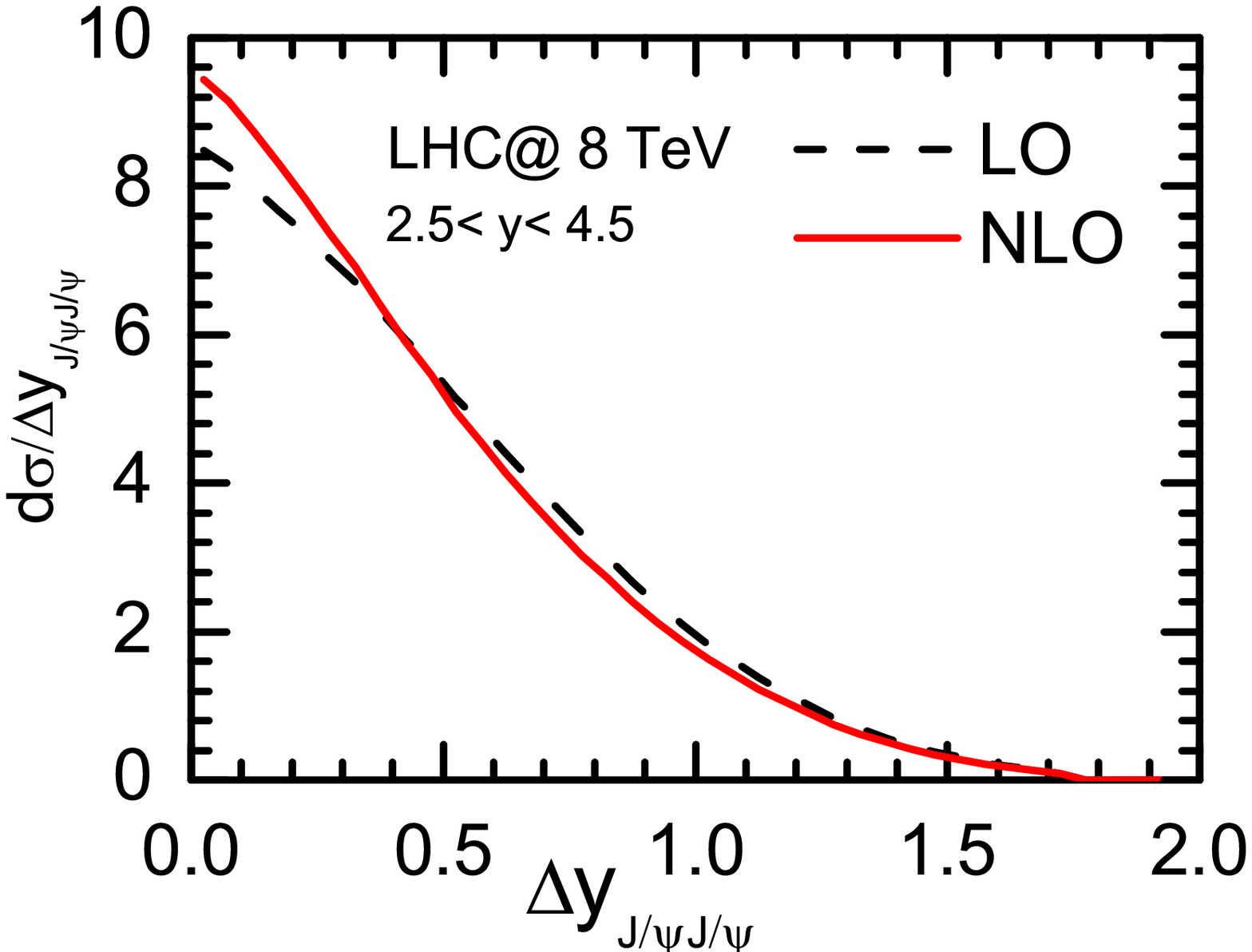}
\caption{\label{fig:detay}  The figure is about the differential cross section to double $J/\psi$ production over rapidities of $J/\psi J/\psi$ at LHC with $\sqrt s=7TeV$ and $2<y<4.5$,and $\sqrt s=8TeV$ and $2.5<y<4.5$ respectively.}
\end{figure}

\begin{figure}[h]
\centering
\includegraphics[width=0.49\textwidth]{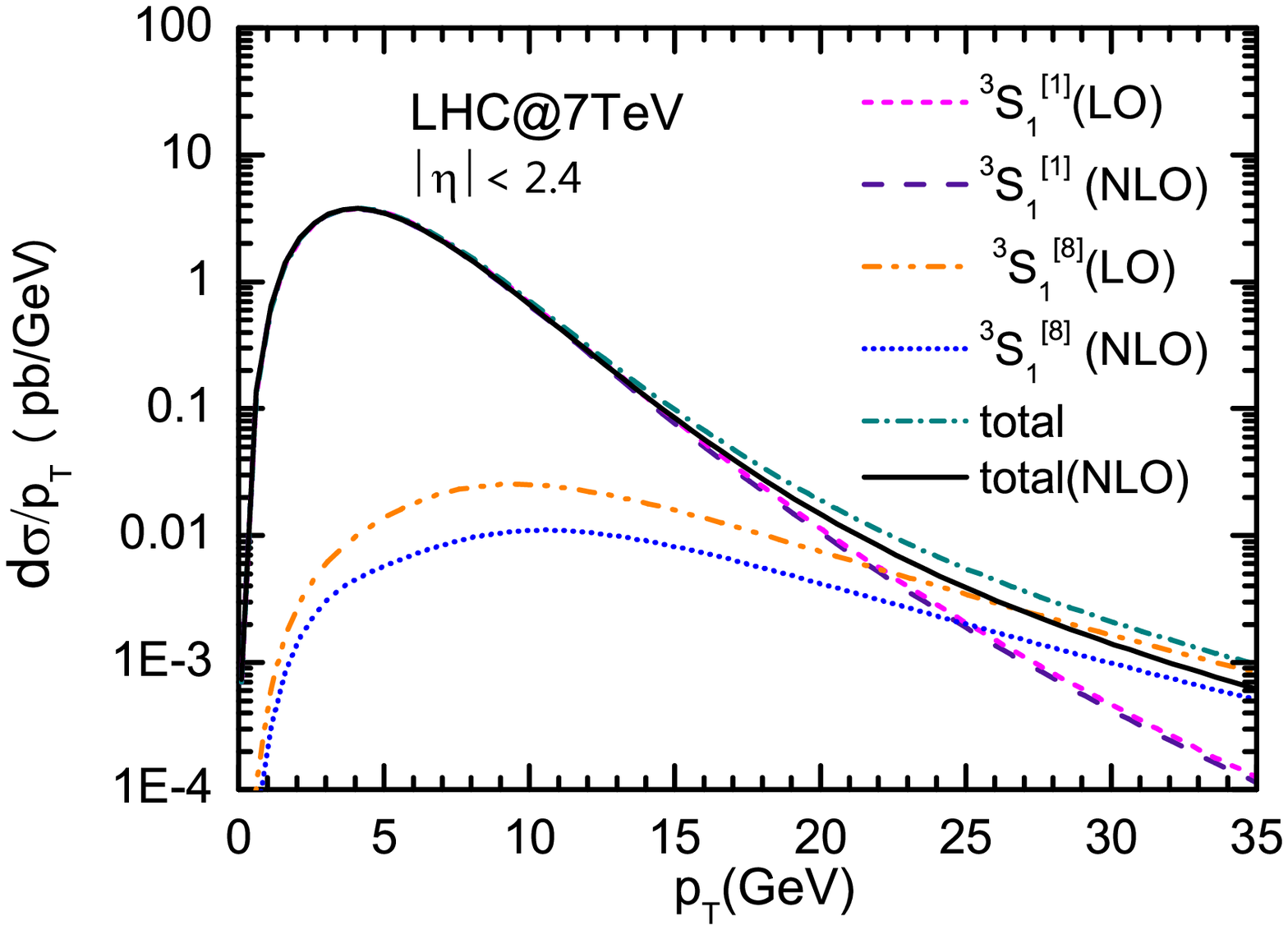}
\includegraphics[width=0.49\textwidth]{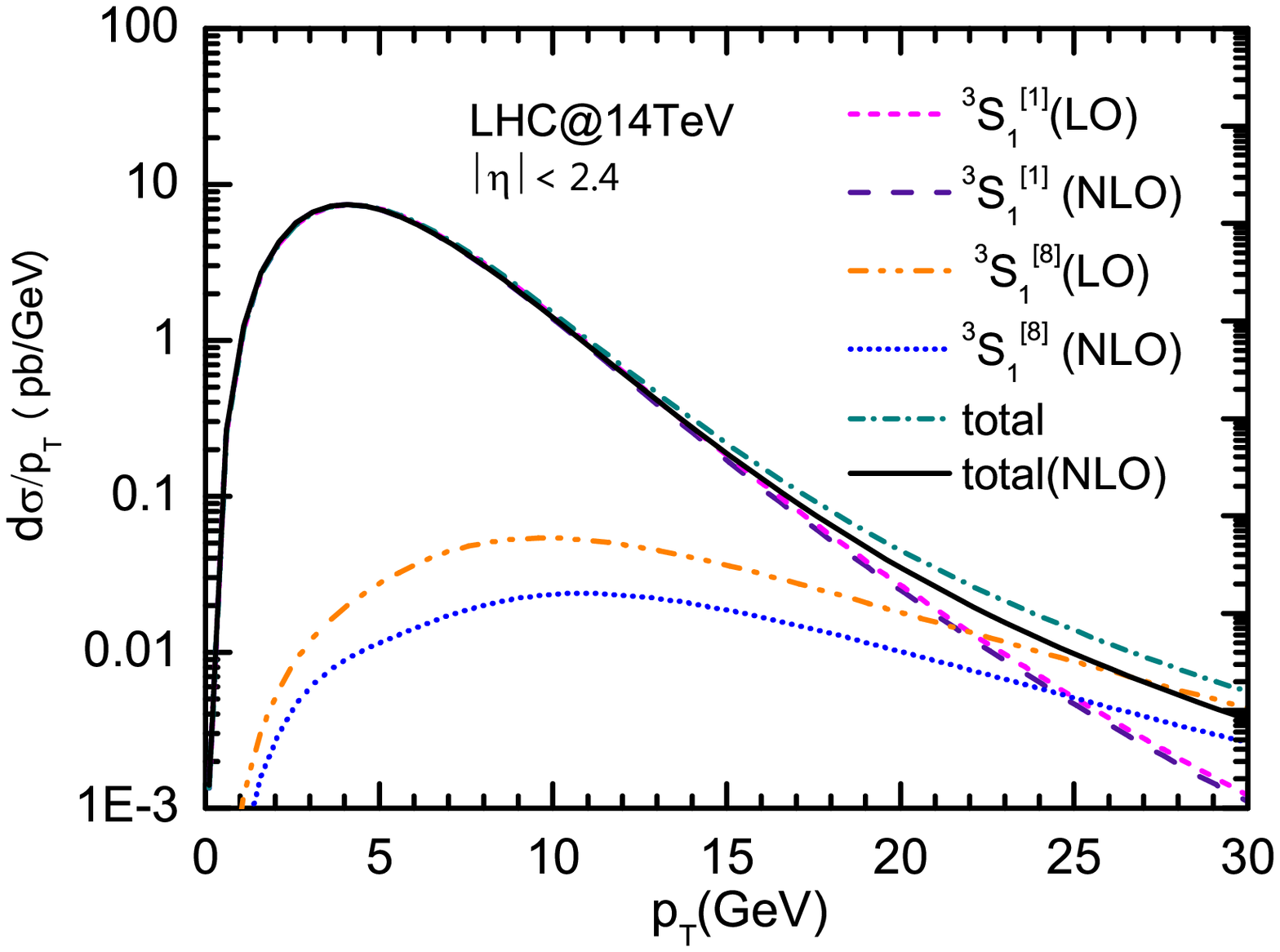}
\caption{\label{fig:bb} The deferential cross section of ${p+p}\rightarrow{\Upsilon+\Upsilon}$ process as a function of $p_t$ at the LHC with $|\eta|<2.4$ and  $\sqrt{s}=7TeV$ , $\sqrt{s}=14TeV$, respectively. }
\end{figure}
\subsection{$J/\psi+\Upsilon$ channel}

We give the  differential cross sections as a function of $p_t$ at the LHC with $|\eta|<2.4$
when choosing the beam energy of the LHC and the LHCb, $\sqrt{s}=7TeV$ and $\sqrt{s}=14TeV$, respectively.

\begin{figure}[h]
\centering
\includegraphics[width=0.49\textwidth]{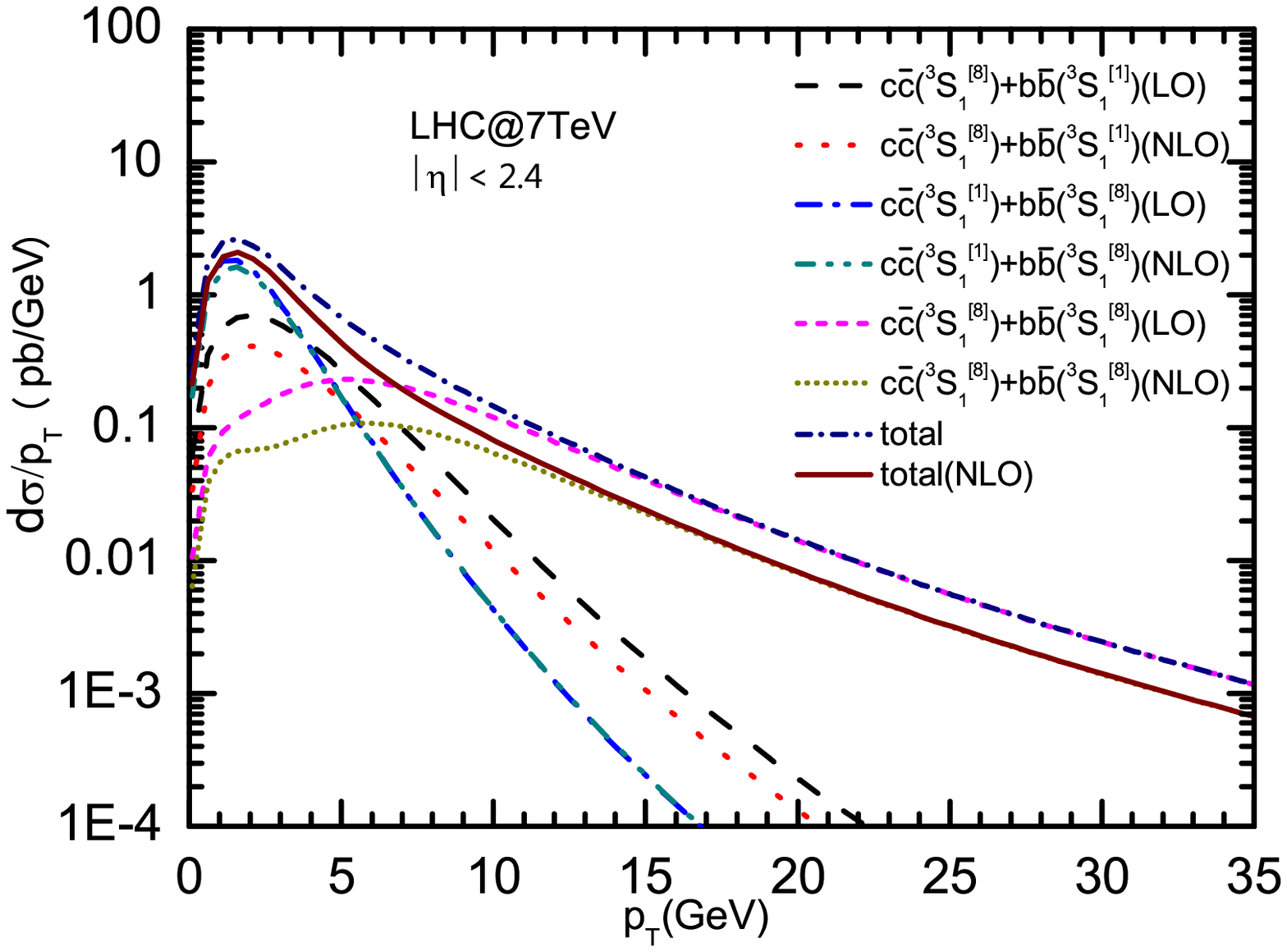}
\includegraphics[width=0.49\textwidth]{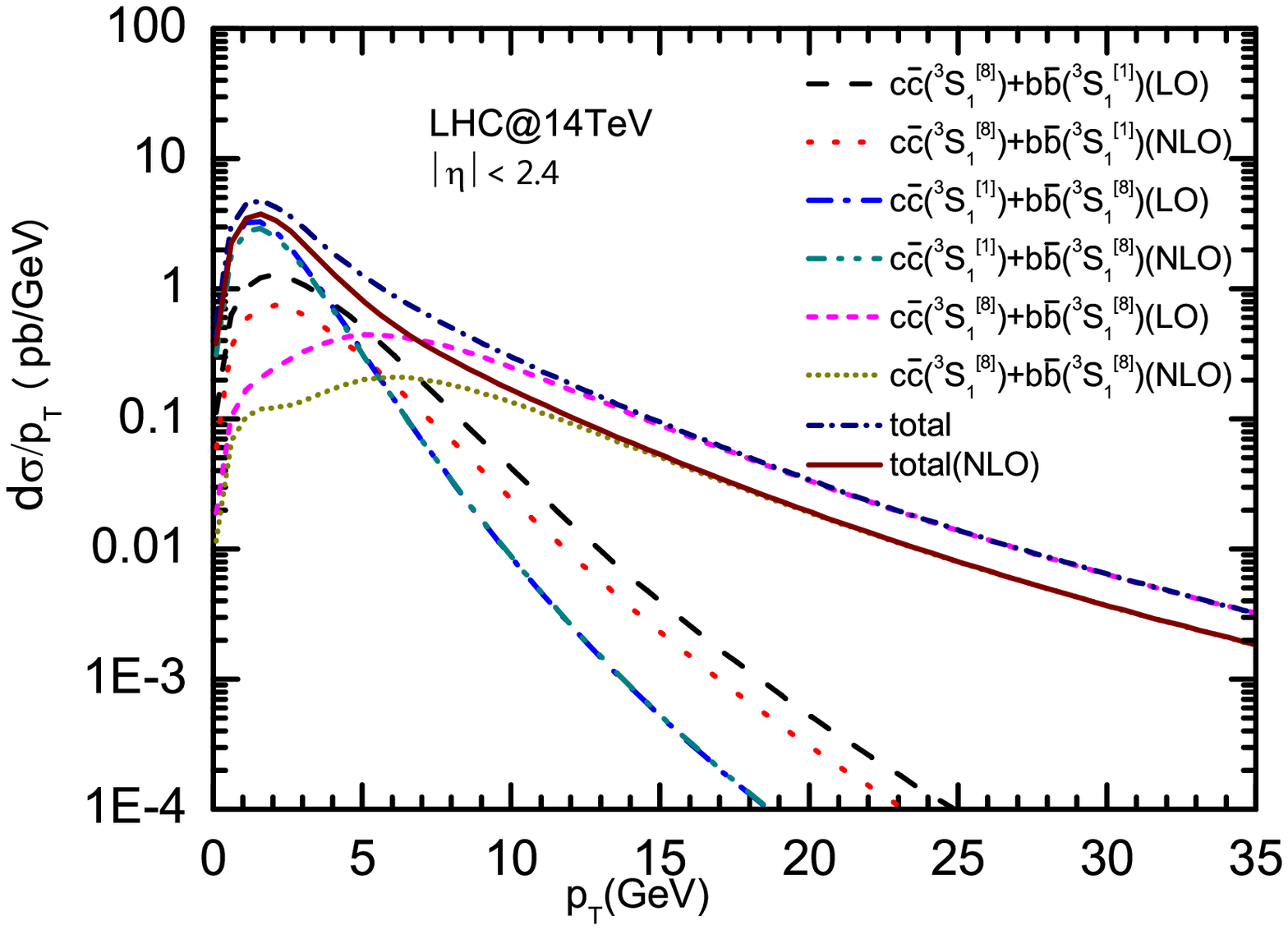}
\caption{\label{fig:ccbb}The differential cross section of ${p+p}\rightarrow{J/\psi+\Upsilon}$ process as a function of $p_t$ at the LHC with $\sqrt{s}=7TeV$ and $\sqrt{s}=14TeV$ with $|\eta|<2.4$ .}
\end{figure}

As discussed in Sec.2 we consider only the relativistic corrections of the $J/\psi$ part.
The radios of $F/G$ in the large $p_T$ limit are given below
\begin{eqnarray}
R[{c\overline{c}}(^3S_{8}^{[8]}),{b\overline{b}}(^3S_{1}^{[8]})]\Big|_{p_T\gg m}=\frac{G[{c\overline{c}}(^3S_1^{[8]}),{b\overline{b}}(^3S_{1}^{[8]})]}{F[{c\overline{c}}(^3S_1^{[8]})
,{b\overline{b}}(^3S_{8}^{[8]})]}\Big|_{p_T\gg m}&\sim&-\frac{11}{6}\nonumber \\
R[{c\overline{c}}(^3S_{8}^{[8]}),{b\overline{b}}(^3S_{1}^{[1]})]\Big|_{p_T\gg m}=\frac{G[{c\overline{c}}(^3S_1^{[8]}),{b\overline{b}}(^3S_{1}^{[1]})]}{F[{c\overline{c}}(^3S_1^{[8]})
,{b\overline{b}}(^3S_{1}^{[1]})]}\Big|_{p_T\gg m}&\sim&-\frac{11}{6}\nonumber \\
\end{eqnarray}
 For the channel ${c\bar{c}}(^3S_{8}^{[1]})+{b\bar{b}}(^3S_{1}^{[8]})$,
 the ratio  is not a const result and its value approximate 1.5 at the large $p_T$ region.
 In Fig.\ref{fig:ccbb}, we give the $P_T$ distribution of the LO cross sections  and the next
 leading order ones for different channels.
The relativistic corrections for the production of $J/\psi+\Upsilon$
are limited to change the LO CO contributions. This process may be still absorbing to research to be a probe to COM.

\section{Summary}

We calculate the relativistic corrections to the process
${p+p(\bar p)}\rightarrow{J/\psi+J/\psi+X}$ ,${p+p(\bar p)}\rightarrow{\Upsilon+\Upsilon+X}$,
and ${p+p(\bar p)}\rightarrow{J/\psi+\Upsilon+X}$ at the Tevatron and at the LHC based on NRQCD.
The ratios of short distance coefficients between relativistic corrections and LO results for CS and CO states
are approximately $-1,-\frac{11}{3}$ in large $p_T$ limit for the $J/\psi$ and $\Upsilon$ pair productions.
And for $J/\psi+\Upsilon$ the ratio is $-\frac{11}{6}$ for the CO channel.

If fix the LO LDMEs and estimated the NLO($v^2$) LDMEs through the velocity scaling rule
with adopting $v^2=0.23$, the differential cross sections are suppressed by $23\%$ and $84\%$
to LO cross sections for CS channel and CO channel at large $p_T$ region, respectively.
The relativistic corrections significantly dilute the difference
between the shape of color-singlet differential cross sections and the color-octet's at leading order.
So, it is difficult to study the color-octet
mechanism in quarkonium production by distinguishing the color-siglet contribution and color-octet contribution at different $p_T$ regime, if the relativistic corrections are considered in pair heavy quarkonia productions at hadron collider.

We also compute the differential cross sections as a function of the double $J/\psi$ invariant mass and
compare it with the recent LHCb experiments. Our results show that the single parton scattering
contributions are large enough to fit experimental data and it would leave limit room to double
parton scattering mechanism.

Note: when this paper is preparing, another two thesises appeared which discussed
the relativistic corrections of the pair production of $J/\psi$ and $\eta_c$ at the hadron collider\cite{Martynenko:2012tf,Martynenko:2013daa}
but they were computed within the relativistic quark model. Comparing with their results,
the corrections under the NRQCD frame are much smaller than that given by relativistic quark model.

\acknowledgments{
The authors would like to thank Professor K.T. Chao for useful
discussion. This work was
supported by the National Natural Science
Foundation of China (Grants No.10805002, No.10875055, and
No.11075011), the Foundation for the Author of National
Excellent Doctoral Dissertation of China (Grants No. 2007B18 and No. 201020), the
Project of Knowledge Innovation Program (PKIP) of Chinese Academy of
Sciences, Grant No. KJCX2.YW.W10, and the Education Ministry of
Liaoning Province.}

\section{Appendix}

\subsection{ short distance coefficients of ${g+g}\rightarrow{J/\psi+J/\psi}$ and ${g+g}\rightarrow{\Upsilon+\Upsilon}$}
The LO short distance coefficients and their relativistic corrections ones for parton-level process ${g(k_1)+g(k_2)}\rightarrow{J/\psi(p_1)+J/\psi(p_2)}$ and ${g(k_1)+g(k_2)}\rightarrow{\Upsilon(p_1)+\Upsilon(p_2)}$ are given in this appendix. Here $m=2 m_c$ or $2 m_b$ represents the $J/\psi$ mass or $\Upsilon$ mass.
(Here we adopt $t(u)$ to represent $t(0)(u(0))$ directly for simplification)

The LO short distance coefficient for CS channel is:
\begin{eqnarray}
&&{F({}^3S_1^{[1]}{}^3S_{1}^{[1]})}=\frac{64\alpha_s^4\pi^3}{6561 m^2 (m^2-t)^4 (m^2-u)^4 (-2 m^2+t+u)^8}\Big[2680 m^{24}-14984 m^{22}(t+u)
\nonumber\\
&&+m^{20}(31406 t^2+89948 t u+31406 u^2)-16 m^{18} (1989 t^3+12661 t^2 u
\nonumber\\
&&+12661 t u^2+1989 u^3)+4 m^{16}(4417 t^4+57140 t^3 u+117714 t^2 u^2+57140 t u^3+4417 u^4)
\nonumber\\
&&-4 m^{14}(1793 t^5+38340 t^4 u+134119 t^3 u^2+134119 t^2 u^3+38340 t u^4+1793 u^5)+m^{12} (2956 t^6
\nonumber\\
&&+76406 t^5 u+361624 t^4u^2+571900 t^3u^3+361624 t^2 u^4+76406 t u^5+2956 u^6)-2 m^{10} (397 t^7
\nonumber\\
&&+15391 t^6u+91227 t^5 u^2+167593 t^4 u^3+167593 t^3 u^4+91227 t^2 u^5+15391 t u^6+397 u^7)
\nonumber\\
&&+m^8 (47 t^8+7642 t^7 u+73146 t^6 u^2+150334 t^5 u^3+132502 t^4u^4+150334 t^3 u^5+73146 t^2 u^6
\nonumber\\
&&+7642 t u^7+47 u^8)+2 m^6 (10 t^9-411t^8 u-8951 t^7 u^2-29063 t^6 u^3-17653 t^5 u^4-17653 t^4 u^5
\nonumber\\
&&-29063 t^3 u^6-8951t^2 u^7-411 t u^8+10 u^9)+m^4 (t^{10}-66 t^9 u+2469 t^8 u^2+12874 t^7u^3
\nonumber\\
&&+11928 t^6 u^4+1164 t^5 u^5+11928 t^4 u^6+12874 t^3 u^7+2469 t^2 u^8-66 t u^9+u^{10})+4 m^2 t^2 u^2 (9 t^7
\nonumber\\
&&-586 t^6 u-37 t^5 u^2-394 t^4 u^3-394t^3 u^4-37 t^2 u^5-586 t u^6+9 u^7)+2 t^4 u^4 (349 t^4-908 t^3 u
\nonumber\\
&&+1374 t^2 u^2-908 t u^3+349 u^4)\Big]
\end{eqnarray}
The LO short distance coefficient for CO channel is:
\begin{eqnarray}
&&{F({}^3S_1^{[8]}{}^3S_{1}^{[8]})}={\frac{\alpha_s^4\pi^3}{972 m^6 \left(m^2-t\right)^4 \left(m^2-u\right)^4 \left(-2
   m^2+t+u\right)^8}}
\nonumber\\
&&\Big[-104624 m^{28}+522544 m^{26} (t+u)-2 m^{24}(137741 t^2+1836350 t u+137741 u^2)
\nonumber\\
&&-16 m^{22} (80955 t^3-345827 t^2 u-345827 t u^2+80955 u^3)+4 m^{20}t(667342 t^4-701017 t^3 u
\nonumber\\
&&-1566126 t^2 u^2-701017 t u^3+667342 u^4)-8 m^{18} (316747 t^5+1677 t^4 u+195158 t^3 u^2
\nonumber\\
&&+195158 t^2 u^3+1677 t u^4+316747 u^5)+m^{16}(1455637 t^6+281021 t^5 u+5910067 t^4 u^2
\nonumber\\
&&+14296294 t^3 u^3+5910067 t^2 u^4+281021 t u^5+1455637 u^6)-2 m^{14} (270334 t^7-179453 t^6 u
\nonumber\\
&&+2260581 t^5 u^2+8407066 t^4 u^3+8407066 t^3 u^4+2260581 t^2 u^5-179453 t u^6+270334 u^7)
\nonumber\\
&&+2 m^{12}(66670 t^8-239050 t^7 u+1059585 t^6 u^2+5792306 t^5 u^3+7997354 t^4 u^4+5792306 t^3 u^5
\nonumber\\
&&+1059585 t^2 u^6-239050 t u^7+66670 u^8)-2 m^{10}(10661 t^9-110448 t^8 u+408662 t^7 u^2
\nonumber\\
&&+3075158 t^6 u^3+5105335 t^5 u^4+5105335 t^4 u^5+3075158 t^3 u^6+408662 t^2 u^7-110448 t u^8
\nonumber\\
&&+10661 u^9)+2 m^8 (935 t^{10}-25638 t^9 u+178326 t^8 u^2+1288892 t^7 u^3+2655522 t^6 u^4
\nonumber\\
&&+3136494 t^5 u^5+2655522 t^4 u^6+1288892 t^3 u^7+178326 t^2 u^8-25638 t u^9+935 u^{10})\nonumber\\
&&+2 m^6 t u (2616 t^9-65577 t^8 u-431588 t^7 u^2-1010108 t^6 u^3-1490351 t^5 u^4-1490351 t^4 u^5
\nonumber\\
&&-1010108 t^3 u^6-431588 t^2 u^7-65577 t u^8+2616 u^9)+m^4 t u (-243 t^{10}+33354 t^9 u
\nonumber\\
&&+192771t^8 u^2+555338 t^7 u^3+879860 t^6 u^4+1154016 t^5 u^5+879860 t^4 u^6+555338 t^3 u^7
\nonumber\\
&&+192771 t^2 u^8+33354 t u^9-243 u^{10})-162 m^2 t^2 u^2 (t+u)^3 (27 t^6+88 t^5 u+174 t^4 u^2
\nonumber\\
&&+208 t^3 u^3+174 t^2 u^4+88 t u^5+27 u^6)+243 t^2 u^2 (t+u)^4 (t^2+t u+u^2)^3\Big]
\end{eqnarray}

The short distance coefficient to relativistic correction for CS channel is:
\begin{eqnarray}
&&{G({}^3S_1^{[1]}{}^3S_{1}^{[1]})}=\frac{256\alpha_s^4\pi^3}{19683 m^4 (m^2-t)^5(m^2-u)^5 (-2 m^2+t+u)^{10}
   (2 m^2+t+u)}
\nonumber\\
&&\Big[370752 m^{34}-2181536 (t+u) m^{32}+64 \big[73726 (t^2+u^2)+211649 u t\big]m^{30}
\nonumber\\
&&-16\big[244805 (t^3+u^3)+1965309 u t(t+u)\big] m^{28}-4\big[316463(t^4+u^4)
\nonumber\\
&&-8160588 u t(t^2+u^2)-18782102 u^2 t^2\big] m^{26}+\big[5600590(t^5+u^5)
\nonumber\\
&&-9749802 u t(t^3+u^3)-79767876 u^2 t^2(t+u)\big] m^{24}-2\big[2766605 (t^6+u^6)
\nonumber\\
&&
+5407076 u t(t^4+u^4)-15844965 u^2 t^2(t^2+u^2)-35085112 u^3 t^3\big] m^{22}
\nonumber\\
&&+2\big[1563020 (t^7+u^7)+6369573 u t(t^5+u^5)+2696929 u^2 t^2(t^3+u^3)
\nonumber\\
&&-456546 u^3t^3(t+u)\big] m^{20}-4\big[295064(t^8+u^8) +1552733 u t(t^6+u^6)
\nonumber\\
&&+1230476 u^2 t^2(t^4+u^4)+7465803 u^3 t^3(t^2+u^2)+17306424 u^4 t^4\big] m^{18}
\nonumber\\
&&+2\big[157467 (t^9+u^9)+858538 u t(t^7+u^7)-1932037 u^2t^2(t^5+u^5)
\nonumber\\
&&+3904293 u^3 t^3(t^3+u^3)+38310315 u^4 t^4(t+u)\big] m^{16}-2 \big[30250 (t^{10}+u^{10})
\nonumber\\
&&+134737 u t(t^8+u^8)-2493940 u^2 t^2(t^6+u^6)-5184812 u^3t^3(t^4+u^4)
\nonumber\\
&&+15398042 u^4 t^4(t^2+u^2)+35880758 u^5 t^5\big] m^{14}+\big[8417 (t^{11}+u^{11})
\nonumber\\
&&+34279 u t(t^9+u^9)-2497569 u^2 t^2(t^7+u^7)-10315939 u^3t^3(t^5+u^5)
\nonumber\\
&&+1057162 u^4 t^4(t^3+u^3)+31198738 u^5 t^5(t+u)\big] m^{12}+\big[-817 (t^{12}+u^{12})
\nonumber\\
&&-13616 u t(t^{10}+u^{10})+672840 u^2t^2(t^8+u^8)+4910632 u^3 t^3(t^6+u^6)
\nonumber\\
&&+5008157 u^4 t^4(t^4+u^4)-7403400 u^5 t^5(t^2+u^2)-13064072 u^6 t^6\big]m^{10}
\nonumber\\
&&-\big[21 (t^{13}+u^{13})-4794 u t(t^{11}+u^{11})+84325 u^2 t^2(t^9+u^9)
\nonumber\\
&&+1346073 u^3 t^3(t^7+u^7)+3171602 u^4t^4(t^5+u^5)-481617 u^5 t^5(t^3+u^3)
\nonumber\\
&&-3414586 u^6 t^6(t+u)\big] m^8+\big[7 (t^{14}+u^{14})-248 ut (t^{12}+u^{12})
\nonumber\\
&&-2777 u^2 t^2(t^{10}+u^{10})+197002 u^3 t^3(t^8+u^8)+973759 u^4 t^4(t^6+u^6)
\nonumber\\
&&+633832 u^5 t^5(t^4+u^4)-734333 u^6t^6(t^2+u^2)-747796 u^7 t^7\big] m^6
\nonumber\\
&&+t u\big[(t^{13}+u^{13})+137 u t(t^{11}+u^{11})-1320 u^2 t^2(t^9+u^9)
\nonumber\\
&&-171964 u^3 t^3(t^7+u^7)-249123 u^4t^4(t^5+u^5)+36693 u^5 t^5(t^3+u^3)
\nonumber\\
&&+125000 u^6 t^6(t+u)\big] m^4+2 t^3 u^3\big[2 (t^{10}+u^{10})+761 u t(t^8+u^8)
\nonumber\\
&&+36064 u^2 t^2(t^6+u^6)-14592 u^3 t^3(t^4+u^4)+9630 u^4 t^4(t^2+u^2)
\nonumber\\
&&-5362 u^5 t^5\big] m^2-2 t^5 u^5\big[227 (t^7+u^7)+6021 ut(t^5+u^5)
\nonumber\\
&&-9217 u^2 t^2(t^3+u^3)+6041 u^3 t^3(t+u)\big]\Big]
\end{eqnarray}

The short distance coefficient to relativistic correction for CO channel is:
\begin{eqnarray}
&&{G({}^3S_1^{[8]}{}^3S_{1}^{[8]})}=\frac{\alpha_s^4\pi^3}{729 m^8 s^2 (m^2-t)^5 (m^2-u)^5 (-2 m^2+t+u)^8 (2 m^2+t+u)}
\nonumber\\
&&\Big[-7840128 m^{38}+58018240 (t+u) m^{36}-16 \big[4351865 (t^2+u^2)+36821254 u t \big] m^{34}
\nonumber\\
&&-8 \big[22135741(t^3+u^3)-183792021 u t(t+u)\big] m^{32}+4\big[173739553 (t^4+u^4)
\nonumber\\
&&-435034596 u t(t^2+u^2) -916724794 u^2t^2\big] m^{30}-2\big[576450337 (t^5+u^5)
\nonumber\\
&&-575408499 u t(t^3+u^3)-2090528238 u^2 t^2(t+u)\big] m^{28}+2\big[612519223 (t^6+u^6)
\nonumber\\
&&-216110060 u t(t^4+u^4) -1597249863 u^2t^2(t^2+u^2)-1708173704 u^3t^3\big] m^{26}
\nonumber\\
&&+2 \big[-462600109 (t^7+u^7)+86546766 u t(t^5+u^5)+1187377312 u^2t^2(t^3+u^3)
\nonumber\\
&&+911415039 u^3 t^3(t+u)\big] m^{24}+\big[519274693(t^8+u^8)-285877238 u t(t^6+u^6)
\nonumber\\
&&-2090011040 u^2 t^2(t^4+u^4)-1710759498 u^3 t^3(t^2+u^2)-1356254154 u^4 t^4\big] m^{22}
\nonumber\\
&&+\big[-221652624 (t^9+u^9)+375759263 u t(t^7+u^7)+1820975245 u^2 t^2(t^5+u^5)
\nonumber\\
&&+1804776207 u^3t^3(t^3+u^3) +2287765845 u^4 t^4(t+u)\big]m^{20}+2\big[36158642 (t^{10}+u^{10})
\nonumber\\
&&
-146055034 u t(t^8+u^8)-613461077 u^2 t^2(t^6+u^6)-576102238 u^3 t^3(t^4+u^4)
\nonumber\\&&
-723890837 u^4 t^4(t^2+u^2)-1150404320 u^5t^5\big] m^{18}+\big[-17656681(t^{11}+u^{11})
\nonumber\\
&&+148210450 u t (t^9+u^9)+547766115 u^2 t^2(t^7+u^7)+168319214 u^3 t^3(t^5+u^5)
\nonumber\\&&
-601363505 u^4 t^4(t^3+u^3)-515358665 u^5 t^5(t+u)\big] m^{16}+2\big[1515472 (t^{12}+u^{12})
\nonumber\\
&&-25773835 u t (t^{10}+u^{10})-63896082 u^2 t^2(t^8+u^8)+200994011 u^3 t^3(t^6+u^6)
\nonumber\\
&&+902427166 u^4 t^4(t^4+u^4)+1542963240 u^5t^5(t^2+u^2)+1808885720 u^6 t^6\big] m^{14}
\nonumber\\
&&-\big[320244 (t^{13}+u^{13})-12358347 u t(t^{11}+u^{11})+10043059 u^2 t^2(t^9+u^9)
\nonumber\\&&
+412260621 u^3 t^3(t^7+u^7)+1621412381u^4 t^4(t^5+u^5)+3279793470 u^5 t^5(t^3+u^3)
\nonumber\\
&&+4552155452 u^6 t^6(t+u)\big] m^{12}+\big[14662 (t^{14}+u^{14})-1965047 u t(t^{12}+u^{12})
\nonumber\\
&&+19157992 u^2t^2(t^{10}+u^{10})+213718612 u^3 t^3(t^8+u^8)+861766390 u^4 t^4(t^6+u^6)
\nonumber\\
&&+1999127983 u^5 t^5(t^4+u^4)+3170423452 u^6 t^6(t^2+u^2)+3695142248 u^7 t^7\big] m^{10}
\nonumber\\
&&+t u \big[190300 (t^{13}+u^{13})-7050997  u t(t^{11}+u^{11})-69336231 u^2 t^2(t^9+u^9)
\nonumber\\&&
-300672607 u^3 t^3(t^7+u^7)-788642811 u^4 t^4(t^5+u^5)-1422124344 u^5 t^5(t^3+u^3)
\nonumber\\
&&-1887768238 u^6 t^6(t+u)\big] m^8+t u\big[-8019(t^{14}+u^{14})+1359759u t(t^{12}+u^{12})
\nonumber\\
&&+14495197 u^2t^2(t^{10}+u^{10})+69109969  u^3 t^3(t^8+u^8)+205737389  u^4 t^4(t^6+u^6)
\nonumber\\
&&+417998865  u^5 t^5(t^4+u^4)+630439593  u^6 t^6(t^2+u^2)+718419166 u^7 t^7\big]m^6
\nonumber\\&&
-t^2 u^2\big[143856  (t^{13}+u^{13})+1824408  u t(t^{11}+u^{11})+10108398 u^2 t^2(t^9+u^9)
\nonumber\\
&&+33845791 u^3 t^3(t^7+u^7)+78611535  u^4 t^4(t^5+u^5)+132750949 u^5 t^5(t^3+u^3)
\nonumber\\
&&+172324471 u^6 t^6(t+u)\big] m^4+81 t^2 u^2 (t+u)^6\big[81 (t^8+u^8)+999 u t(t^6+u^6)
\nonumber\\
&&+2910 u^2 t^2(t^4+u^4)+5459 u^3 t^3(t^2+u^2)+6452u^4 t^4\big] m^2
\nonumber\\
&&-2673 t^3 u^3 (t+u)^7 (t^2+u t+u^2)^3\Big]
\end{eqnarray}
\subsection{ short distance coefficients of ${g+g}\rightarrow{J/\psi+\Upsilon}$}

The LO short distance coefficient for ${g+g}\rightarrow{c\overline{c}(^3S_1^{[8]})+b\overline{b}(^3S_1^{[1]})}$ channel is:
\begin{eqnarray}
&&{F({c\overline{c}}(^3S_1^{[8]}){b\overline{b}}(^3S_{1}^{[1]}))}=\frac{10 \alpha_s^4\pi^3}
{243 m^3{M} ({M}^2-t)^2({M}^2-u)^2 (-2m^2+t+u)^2  s^2}
\nonumber\\
&&\Big[2 m^8 {M}^2+2m^6 (7 {M}^4-5 {M}^2 (t+u)+t u)+2 m^4 (7 {M}^6-13{M}^4 (t+u)
\nonumber\\
&&+{M}^2 (7 t^2+10t u+7 u^2)-t u (t+u))+m^2 ({M}^8-8 {M}^6(t+u)
\nonumber\\
&&+{M}^4(15 t^2+16 t u+15u^2)-4 {M}^2 (2 t^3+3 t^2 u+3 t u^2+2 u^3)+t^2 u^2)
\nonumber\\
&&+2 {M}^2 ({M}^4 (t^2+tu+u^2)-{M}^2 (2 t^3+3 t^2 u+3 t u^2+2 u^3)
\nonumber\\
&&+(t^2+t u+u^2)^2)\Big],
\end{eqnarray}
where $m=m_{J/\psi}$, and $M=m_\Upsilon$.

The relativistic of  short distance coefficient for ${g+g}\rightarrow{c\overline{c}(^3S_1^{[8]})+b\overline{b}(^3S_1^{[1]})}$ channel is:
\begin{eqnarray}
&&{G({c\overline{c}}(^3S_1^{[8]}){b\overline{b}}(^3S_{1}^{[1]}))}=-20 \alpha_s^4\pi^3/(729m^5Ms^2({M}^2-t)^3({M}^2-u)^3(4 m^2{M}^2-(t+u)^2))
\nonumber\\
&&/(-2m^2+t+u)^3 \times
\bigg[ 96{M}^4(2{M}^2-t-u)m^{14}+16{M}^2
   \big[61{M}^6-82 (t+u){M}^4+(30 (t^2+u^2)
   \nonumber\\
&&+37 u t){M}^2+3 t u (t+u)\big]m^{12}+4\big[52{M}^{10}-374 (t+u){M}^8+(549(t^2+u^2)+290 u t){M}^6
\nonumber\\
&&+(-227 (t^3+u^3)+61 u t(u+ t)){M}^4-t u (85 (t^2+u^2)+214 u t){M}^2+12 t^2 u^2 (t+u)\big]m^{10}
\nonumber\\
&&-2\big[424 {M}^{12}-916 (t+u){M}^{10}+2(77(t^2+u^2)+934 u t){M}^8+(867 (t^3+u^3)
\nonumber\\
&&-1439 u t(t+u)){M}^6+( -505 (t^4+u^4)+728 u t(t^2+u^2)+1778 u^2 t^2) {M}^4-t u   (241 (t^3+u^3)\nonumber\\
&&+735 u t(u+t)){M}^2+26 t^2 u^2 (t+u)^2\big]m^8-2\big[44{M}^{14}-512 (t+u){M}^{12}+2(545 (t^2+u^2)
\nonumber\\
&&+972 u t){M}^{10}-(451 (t^3+u^3)+2589 u t(t+u)){M}^8+(-572 (t^4+u^4)+1158 u t(t^2+u^2)
\nonumber\\
&&+2264 u^2 t^2){M}^6+(401 (t^5+u^5)-155 u t(t^3+u^3)-1018 u^2 t^2(t+u)){M}^4+2 t u (41 (t^4+u^4)
\nonumber\\
&&+128 u t(t^2+u^2)+196 u^2 t^2) {M}^2-3t^2 u^2 (t+u)^3\big]m^6
+2\big[10 (t+u){M}^{14}-(239 (t^2+u^2)\nonumber\\
&&+390 u t){M}^{12}+2(252 (t^3+u^3)+611 u t(t+u)){M}^{10}-(117 (t^4+u^4)+1676 u^2 t^2\nonumber\\
&&+1027 ut(u^2+ t^2)){M}^8-(383 (t^5+u^5)+244 u t(t^3+u^3)-157 u^2 t^2(t+u)) {M}^6\nonumber\\
&&+(225 (t^6+u^6)+485 u t(t^4+u^4)+670 u^2 t^2(t^2+u^2)+684 u^3 t^3){M}^4
-t u(56 (t^5+u^5)\nonumber\\
&&+205 u t(t^3+u^3)+325 u^2 t^2(t+u)){M}^2-t^2 u^2 (t+u)^2(t^2-9 u t+u^2)\big]m^4+\big[(71 (t^3+u^3)
\nonumber\\
&&+173 u t(t+u)) {M}^{12}-(t+u)^2
   (79 (t^2+u^2)+90 u t){M}^{10}-(211 (t^5+u^5)+590 u t(t^3+u^3)
   \nonumber\\
&&+859 u^2 t^2(t+u)){M}^8+(t+u)^2(375 (t^4+u^4)+634 u t(t^2+u^2)+1006 u^2 t^2){M}^6
\nonumber\\
&&-(156 (t^7+u^7)+827 u t(t^5+u^5)+2082 u^2 t^2(t^3+u^3)+3115 u^3 t^3(t+u)
){M}^4\nonumber\\
&&+t u (t+u)^2 (108  (t^4+u^4)+241 u t(t^2+u^2)+334u^2 t^2){M}^2-5 t^3 u^3 (t+u)^3\big]m^2
\nonumber\\
&&-22{M}^2({M}^2-t)({M}^2-u) (t+u)^3\big[(t^2+ut+u^2){M}^4-(2 (t^3+u^3)+3 u t(t+u)){M}^2\nonumber\\
&&+(t^2+u
   t+u^2)^2\big]\bigg]
\end{eqnarray}

The short distance coefficient of
${gg}\rightarrow{c\overline{c}(^3S_1^{[1]})+b\overline{b}(^3S_1^{[8]})}$ can be get through Exchanging $M$ and $m$.
The relativistic correction of this cross section is :
\begin{eqnarray}
&&{G({c\overline{c}}(^3S_1^{[1]}){b\overline{b}}(^3S_{1}^{[8]}))}=\frac{ 20 \alpha_s^4\pi^3/(729 m^3 {M}^3 s^2)}{(m^2-t)^3 (m^2-u)^3(-2 {M}^2+t+u)^4(4m^2{M}^2-(t+u)^2)}
\nonumber\\
&&\Big[(16 {M}^4 \big[14{M}^2-9 (t+u)\big]m^{16}+4{M}^2\big[300{M}^6-452 (t+u){M}^4+(161 (t^2+u^2)+274 u t){M}^2
\nonumber\\
&&-9 (t^3+ u^3)+13 t u(t+u)\big]m^{14}-\big[288
   {M}^{10}+1648 (t+u){M}^8-4 (575 (t^2+u^2)+894 u t) {M}^6
   \nonumber\\
&&+16(32 (t^3+u^3)+3 u t(t+u)){M}^4+(t+u)^2(3 (t^2+u^2)+574 u t){M}^2-4 (t+u)^3(3 (t^2+u^2)
   \nonumber\\
&&+2 u t)\big]
   m^{12}-\big[800{M}^{12}-2080 (t+u){M}^{10}+16(39( t^2+u^2)+73 u t){M}^8+4(121(t^3+u^3)
   \nonumber\\
&&-65 u t(t+u)){M}^6+(34(t^4+u^4)+2916 u t(t^2+u^2)+5444 u^2 t^2){M}^4+(17 (t^5+u^5)
\nonumber\\
&&-807 u t(t^3+u^3)-2346 u^2 t^2(t+u)){M}^2+2 (t+u)^4 (5 ( t^2+u^2)-u t)\big]m^{10}\nonumber\\
&&+\big[
-768 (t+u){M}^{12}+800{M}^{14}-48(t^2+32 u t+u^2){M}^{10}-48(19 (t^3+u^3)-31 t u(t+u)){M}^8
\nonumber\\
&&+
2(787 (t^4+u^4)-122 u t(t^2+u^2)-1546 u^2 t^2){M}^6+(-932(t^5+u^5)+688 u t(t^3+u^3)
   \nonumber\\
&&+6004 u^2 t^2(t+u))
   {M}^4+(t+u)^2(347 (t^4+u^4)-133 u t(t^2+u^2)-592 u^2 t^2){M}^2
  \nonumber\\
&&-2 (t+u)^3(15 (t^4+u^4)+20 u t(t^2+u^2)+26 u^2 t^2)\big]m^8+\big[256 {M}^{16}-992(t+u){M}^{14}
\nonumber\\
&&+8(209 (t^2+u^2)+34 u t){M}^{12}-48(55 (t^3+u^3)-29 u t(t+u)) {M}^{10}+2(1841 (t^4+u^4)
\nonumber\\
&&+704 u t(t^2+u^2)-1722 u^2 t^2) {M}^8-8 (395 (t^5+u^5)+478 u t (t^3+u^3)-183 u^2 t^2(t+u)) {M}^6  \nonumber\\
&&+2 (t+u)^2(782(t^4+u^4)
 -808 u^2 t^2-181  u t(t^2+u^2)) {M}^4-(435 (t^7+u^7)+1327 u  t(t^5+u^5)
\nonumber\\
&&+1381 u^2 t^2(t^3+u^3)
+617 u^3 t^3(t+u)+28 u^2 t^2){M}^2+2(t+u)^4(21 (t^4+u^4)
\nonumber\\
&&+6 u t(t^2+u^2))\big]m^6+\big[32(t^2-13 u t+u^2){M}^{14}-8(41 (t^3+u^3)-229 u t(t+u)){M}^{12}
\nonumber\\
&&+8 (129 (t^4+u^4)-249 u t(t^2+u^2)-644 u^2 t^2
) {M}^{10}-2(765 (t^5+u^5)+85 u t(t^3+u^3)\nonumber\\
&&-2282 u^2 t^2(t+u)){M}^8+2 (t+u)^2(617 (t^4+u^4)
-644 u^2 t^2-484u t(t^2+u^2)){M}^6
\nonumber\\
&&-2 (285 (t^7+u^7)
+559  u  t(t^5+u^5)-315 u^2 t^2(t^3+u^3)-1753 u^3 t^3(t+u)){M}^4
   \nonumber\\
&&+(t+u)^2(144 (t^6+u^6)+159 u t (t^4+u^4)-129 u^2 t^2(t^2+u^2)-280 u^3 t^3){M}^2
\nonumber\\
&&-2 (t+u)^3(7(t^6+u^6)+6 u t (t^4+u^4)- u^2 t^2(t^2+u^2)-8 u^3 t^3)\big]m^4
\nonumber\\
&&-t u \big[-8(9 t^2-10 u t+9 u^2){M}^{12}+8(31 (t^3+u^3)+25 u t(t+u)){M}^{10}-2(141 (t^4+u^4)
\nonumber\\
&&+328 u t(t^2+u^2)+334 u^2 t^2){M}^8+4 (25 (t^5+u^5)+84 u t(t^3+u^3)+107 u^2 t^2(t+u)){M}^6
\nonumber\\
&&+4 (t+u)^2 (3 (t^4+u^4)+20  u t(t^2+u^2)+35 u^2 t^2){M}^4-(t+u)^3(8 (t^4+u^4)
\nonumber\\
&&+27  u t(t^2+u^2)+34 u^2 t^2){M}^2+2 (t+u)^4 (t^2+u t+u^2)^2\big]  m^2\nonumber\\
&&+{M}^2 t^2 u^2 (t+u)^2 (8 {M}^8-16 (t+u) {M}^6+2 (t^2
+12 u t+u^2) {M}^4
\nonumber\\
&&+(6 (t^3+u^3)-2 u t(t+u))
   {M}^2+t u (t+u)^2))\Big]\nonumber\\
\end{eqnarray}
For the process ${g+g}\rightarrow{c\overline{c}(^3S_1^{[8]})+b\overline{b}(^3S_1^{[8]})}$, the express of relativistic correction term is too long to give here. We only give the LO cross section.
\begin{eqnarray}
&&{F({c\overline{c}}(^3S_1^{[8]}){b\overline{b}}(^3S_{1}^{[8]}))}={\alpha_s^4\pi^3}
/(108m^3{M}^3s^2)/\Big[{(m^2\hspace{-0.049cm}-\hspace{-0.049cm}t)}(m^2\hspace{-0.049cm}-\hspace{-0.049cm}u)({M}^2\hspace{-0.049cm}-\hspace{-0.049cm}t)({M}^2\hspace{-0.049cm}-\hspace{-0.049cm}u)\Big]^2\times
\nonumber\\
&&\hspace{-0.4cm}1/\Big[(t\hspace{-0.049cm}+\hspace{-0.049cm}u
\hspace{-0.049cm}-\hspace{-0.049cm}2m^2)(t\hspace{-0.049cm}+\hspace{-0.049cm}u
\hspace{-0.049cm}-\hspace{-0.049cm}2{M}^2)\Big]^2\hspace{-0.1cm}
\times\hspace{-0.1cm}\Big[92{M}^4(t\hspace{-0.049cm}+\hspace{-0.049cm}u\hspace{-0.049cm}-\hspace{-0.049cm}2{M}^2)^2
m^{16}\hspace{-0.049cm}+\hspace{-0.049cm}4{M}^2(767{M}^8\hspace{-0.049cm}-\hspace{-0.049cm}1450(t\hspace{-0.049cm}+\hspace{-0.049cm}u){M}^6\nonumber\\
&&\hspace{-0.4cm}+18(49(t^2\hspace{-0.049cm}+\hspace{-0.049cm}u^2)\hspace{-0.049cm}+\hspace{-0.049cm}99ut){M}^4
\hspace{-0.049cm}-\hspace{-0.049cm}2(86(t^3\hspace{-0.049cm}+\hspace{-0.049cm}u^3)\hspace{-0.049cm}+\hspace{-0.049cm}279ut(t\hspace{-0.049cm}+\hspace{-0.049cm}u)){M}^2\hspace{-0.049cm}+\hspace{-0.049cm}tu(33ut\hspace{-0.049cm}+\hspace{-0.049cm}7(t^2\hspace{-0.049cm}+\hspace{-0.049cm}u^2)))m^{14}\nonumber\\
&&\hspace{-0.4cm}+2(2896{M}^{12}\hspace{-0.049cm}-\hspace{-0.049cm}9812(t\hspace{-0.049cm}+\hspace{-0.049cm}u){M}^{10}\hspace{-0.049cm}+\hspace{-0.049cm}4(5417ut
\hspace{-0.049cm}+\hspace{-0.049cm}2899(t^2\hspace{-0.049cm}+\hspace{-0.049cm}u^2){M}^8\hspace{-0.049cm}-\hspace{-0.049cm}2(2847(t^3\hspace{-0.049cm}+\hspace{-0.049cm}u^3)\nonumber\\
&&\hspace{-0.4cm}+7877ut(t\hspace{-0.049cm}+\hspace{-0.049cm}u)){M}^6+(1003(t^4+u^4)
\hspace{-0.049cm}+\hspace{-0.049cm}6030u^2t^2\hspace{-0.049cm}+\hspace{-0.049cm}3710ut(u^2\hspace{-0.049cm}+\hspace{-0.049cm}t^2)){M}^4\hspace{-0.049cm}-\hspace{-0.049cm}2tu(19(t^3\hspace{-0.049cm}+\hspace{-0.049cm}u^3)\nonumber\\
&&\hspace{-0.4cm}+211ut(t\hspace{-0.049cm}+\hspace{-0.049cm}u)){M}^2\hspace{-0.049cm}+\hspace{-0.049cm}76t^2u^2(t^2\hspace{-0.049cm}+\hspace{-0.049cm}ut
\hspace{-0.049cm}+\hspace{-0.049cm}u^2))m^{12}\hspace{-0.049cm}+\hspace{-0.049cm}(3068{M}^{14}\hspace{-0.049cm}-\hspace{-0.049cm}19624(t\hspace{-0.049cm}+\hspace{-0.049cm}u){M}^{12}\nonumber\\
&&\hspace{-0.4cm}+(41811(u^2\hspace{-0.049cm}+\hspace{-0.049cm}t^2)\hspace{-0.049cm}+\hspace{-0.049cm}75706ut){M}^{10}\hspace{-0.049cm}-\hspace{-0.049cm}9(4421(t^3\hspace{-0.049cm}+\hspace{-0.049cm}u^3)
\hspace{-0.049cm}+\hspace{-0.049cm}11211ut(t\hspace{-0.049cm}+\hspace{-0.049cm}u)){M}^8\hspace{-0.049cm}+\hspace{-0.049cm}2(8729(t^4\hspace{-0.049cm}+\hspace{-0.049cm}u^4)\nonumber\\
&&\hspace{-0.4cm}+27487ut(t^2\hspace{-0.049cm}+\hspace{-0.049cm}u^2)\hspace{-0.049cm}+\hspace{-0.049cm}
40854u^2t^2){M}^6
\hspace{-0.049cm}-\hspace{-0.049cm}
(2924(t^5\hspace{-0.049cm}+\hspace{-0.049cm}u^5)\hspace{-0.049cm}+\hspace{-0.049cm}
10065ut(t^3\hspace{-0.049cm}+\hspace{-0.049cm}
u^3)\hspace{-0.049cm}+\hspace{-0.049cm}23123u^2t^2(t\hspace{-0.049cm}+\hspace{-0.049cm}u)){M}^4\nonumber\\
&&\hspace{-0.4cm}+tu(2303ut(t^2\hspace{-0.049cm}+\hspace{-0.049cm}u^2)\hspace{-0.049cm}-\hspace{-0.049cm}188(t^4\hspace{-0.049cm}+\hspace{-0.049cm}u^4)\hspace{-0.049cm}+\hspace{-0.049cm}3378u^2t^2
){M}^2\hspace{-0.049cm}-\hspace{-0.049cm}8t^2u^2(84(t^3\hspace{-0.049cm}+\hspace{-0.049cm}u^3)\hspace{-0.049cm}+\hspace{-0.049cm}149ut(u\hspace{-0.049cm}+\hspace{-0.049cm}t)))m^{10}
\nonumber\\
&&\hspace{-0.4cm}+(368{M}^{16}\hspace{-0.049cm}-\hspace{-0.049cm}5800(t\hspace{-0.049cm}+\hspace{-0.049cm}u){M}^{14}
\hspace{-0.049cm}+\hspace{-0.049cm}8(2899(t^2\hspace{-0.049cm}+\hspace{-0.049cm}u^2)\hspace{-0.049cm}+\hspace{-0.049cm}5417ut){M}^{12}\hspace{-0.049cm}-\hspace{-0.049cm}9(4421(t^3\hspace{-0.049cm}+\hspace{-0.049cm}u^3)\nonumber\\
&&\hspace{-0.4cm}+11211ut(t\hspace{-0.049cm}+\hspace{-0.049cm}u)){M}^{10}\hspace{-0.049cm}+\hspace{-0.049cm}
(33691
(t^4\hspace{-0.049cm}+\hspace{-0.049cm}u^4)
\hspace{-0.049cm}+\hspace{-0.049cm}100610ut(t^2\hspace{-0.049cm}+\hspace{-0.049cm}u^2)
\hspace{-0.049cm}+\hspace{-0.049cm}147662u^2t^2){M}^8\hspace{-0.049cm}-\hspace{-0.049cm}
2(6949(t^5\hspace{-0.049cm}
\hspace{-0.049cm}+\hspace{-0.049cm}u^5)\nonumber\\
&&\hspace{-0.4cm}+21495ut(t^3\hspace{-0.049cm}+\hspace{-0.049cm}u^3)\hspace{-0.049cm}+\hspace{-0.049cm}43712u^2t^2(t\hspace{-0.049cm}+\hspace{-0.049cm}u)
){M}^6\hspace{-0.049cm}+\hspace{-0.049cm}(2274(t^6\hspace{-0.049cm}+\hspace{-0.049cm}u^6)\hspace{-0.049cm}+\hspace{-0.049cm}4949ut(t^4\hspace{-0.049cm}+\hspace{-0.049cm}u^4)\hspace{-0.049cm}+\hspace{-0.049cm}29902u^3t^3\nonumber\\
&&\hspace{-0.4cm}+20878u^2t^2(t^2\hspace{-0.049cm}+\hspace{-0.049cm}u^2)
){M}^4\hspace{-0.049cm}+\hspace{-0.049cm}tu(718(t^5\hspace{-0.049cm}+\hspace{-0.049cm}u^5)\hspace{-0.049cm}-\hspace{-0.049cm}4033ut(t^3\hspace{-0.049cm}+\hspace{-0.049cm}u^3)\hspace{-0.049cm}-\hspace{-0.049cm}8973u^2t^2(t\hspace{-0.049cm}+\hspace{-0.049cm}u)){M}^2\nonumber\\
&&\hspace{-0.4cm}+2t^2u^2(679(t^4\hspace{-0.049cm}+\hspace{-0.049cm}u^4)
\hspace{-0.049cm}+\hspace{-0.049cm}1669ut(t^2\hspace{-0.049cm}+\hspace{-0.049cm}u^2)\hspace{-0.049cm}+\hspace{-0.049cm}2164u^2t^2))m^8\hspace{-0.049cm}-\hspace{-0.049cm}(368(t\hspace{-0.049cm}+\hspace{-0.049cm}u)
{M}^{16}\hspace{-0.049cm}-\hspace{-0.049cm}72(49(t^2\hspace{-0.049cm}+\hspace{-0.049cm}u^2)\nonumber\\
&&\hspace{-0.4cm}+99ut){M}^{14}\hspace{-0.049cm}+\hspace{-0.049cm}4(2847(t^3\hspace{-0.049cm}+\hspace{-0.049cm}u^3)\hspace{-0.049cm}+\hspace{-0.049cm}7877ut(t\hspace{-0.049cm}+\hspace{-0.049cm}u)){M}^{12}
\hspace{-0.049cm}-\hspace{-0.049cm}2(8729(t^4\hspace{-0.049cm}+\hspace{-0.049cm}u^4)\hspace{-0.049cm}+\hspace{-0.049cm}27487ut(t^2\hspace{-0.049cm}+\hspace{-0.049cm}u^2)
\nonumber\\
&&\hspace{-0.4cm}+40854u^2t^2){M}^{10}\hspace{-0.049cm}+\hspace{-0.049cm}2(6949(t^5\hspace{-0.049cm}+\hspace{-0.049cm}u^5)
\hspace{-0.049cm}+\hspace{-0.049cm}21495ut(t^3\hspace{-0.049cm}+\hspace{-0.049cm}u^3)\hspace{-0.049cm}+\hspace{-0.049cm}43712u^2
t^2(t\hspace{-0.049cm}+\hspace{-0.049cm}u)){M}^8\hspace{-0.049cm}-\hspace{-0.049cm}
(5580(t^6\hspace{-0.049cm}+\hspace{-0.049cm}u^6)\nonumber\\
&&\hspace{-0.4cm}+12481ut(t^4\hspace{-0.049cm}+\hspace{-0.049cm}u^4)\hspace{-0.049cm}+\hspace{-0.049cm}42144u^2t^2(t^2\hspace{-0.049cm}+\hspace{-0.049cm}u^2)\hspace{-0.049cm}+\hspace{-0.049cm}58166u^3t^3){M}^6
\hspace{-0.049cm}+\hspace{-0.049cm}(912(t^7\hspace{-0.049cm}+\hspace{-0.049cm}u^7)\hspace{-0.049cm}-\hspace{-0.049cm}893ut(t^5\hspace{-0.049cm}+\hspace{-0.049cm}u^5)\nonumber\\
&&\hspace{-0.4cm}
\hspace{-0.049cm}+\hspace{-0.049cm}11979u^2t^2(t^3\hspace{-0.049cm}+\hspace{-0.049cm}u^3)
\hspace{-0.049cm}+\hspace{-0.049cm}26506u^3t^3(t\hspace{-0.049cm}+\hspace{-0.049cm}u)){M}^4
\hspace{-0.049cm}+\hspace{-0.049cm}tu(762(t^6\hspace{-0.049cm}+\hspace{-0.049cm}u^6)
\hspace{-0.049cm}-\hspace{-0.049cm}5413ut(t^4\hspace{-0.049cm}+\hspace{-0.049cm}u^4)
\hspace{-0.049cm}-\hspace{-0.049cm}21038u^3t^3\nonumber\\
&&\hspace{-0.4cm}-15468u^2t^2(t^2\hspace{-0.049cm}+\hspace{-0.049cm}u^2)
){M}^2\hspace{-0.049cm}+\hspace{-0.049cm}2t^2u^2
(789(t^5\hspace{-0.049cm}+\hspace{-0.049cm}u^5)\hspace{-0.049cm}+\hspace{-0.049cm}2413ut
(t^3\hspace{-0.049cm}+\hspace{-0.049cm}u^3)
\hspace{-0.4cm}+\hspace{-0.049cm}3910u^2t^2(t\hspace{-0.049cm}+\hspace{-0.049cm}u)))m^6
\nonumber\\
&&\hspace{-0.4cm}+(92(t\hspace{-0.049cm}+\hspace{-0.049cm}u)^2{M}^{16}\hspace{-0.049cm}-\hspace{-0.049cm}8(86(t^3\hspace{-0.049cm}+\hspace{-0.049cm}u^3)\hspace{-0.049cm}+\hspace{-0.049cm}279ut(t\hspace{-0.049cm}+\hspace{-0.049cm}u)3){M}^{14}\hspace{-0.049cm}+\hspace{-0.049cm}2(1003(t^4\hspace{-0.049cm}+\hspace{-0.049cm}u^4)\hspace{-0.049cm}+\hspace{-0.049cm}6030u^2t^2
\nonumber\\
&&\hspace{-0.4cm}+3710ut(t^2\hspace{-0.049cm}+\hspace{-0.049cm}u^2)){M}^{12}\hspace{-0.049cm}-\hspace{-0.049cm}(2924(t^5\hspace{-0.049cm}+\hspace{-0.049cm}u^5)\hspace{-0.049cm}+\hspace{-0.049cm}10065ut(t^3\hspace{-0.049cm}+\hspace{-0.049cm}u^3)\hspace{-0.049cm}+\hspace{-0.049cm}23123u^2t^2(t\hspace{-0.049cm}+\hspace{-0.049cm}u)){M}^{10}
\nonumber\\
&&\hspace{-0.4cm}+(2274(t^6\hspace{-0.049cm}+\hspace{-0.049cm}u^6)\hspace{-0.049cm}+\hspace{-0.049cm}4949ut(t^4\hspace{-0.049cm}+\hspace{-0.049cm}u^4)\hspace{-0.049cm}+\hspace{-0.049cm}20878u^2t^2(t^2\hspace{-0.049cm}+\hspace{-0.049cm}u^2)\hspace{-0.049cm}+\hspace{-0.049cm}29902u^3t^3){M}^8\hspace{-0.049cm}-\hspace{-0.049cm}(912(t^7\hspace{-0.049cm}+\hspace{-0.049cm}u^7)
\nonumber\\
&&\hspace{-0.4cm}-893ut(t^5\hspace{-0.049cm}+\hspace{-0.049cm}u^5)\hspace{-0.049cm}+\hspace{-0.049cm}11979u^2t^2(t^3\hspace{-0.049cm}+\hspace{-0.049cm}u^3)\hspace{-0.049cm}+\hspace{-0.049cm}26506u^3t^2(t\hspace{-0.049cm}+\hspace{-0.049cm}u)){M}^6\hspace{-0.049cm}+\hspace{-0.049cm}(152(t^8\hspace{-0.049cm}+\hspace{-0.049cm}u^8)
\nonumber\\
&&\hspace{-0.4cm}-1483ut(t^6\hspace{-0.049cm}+\hspace{-0.049cm}u^6)\hspace{-0.049cm}+\hspace{-0.049cm}8243u^2t^2(t^4\hspace{-0.049cm}+\hspace{-0.049cm}u^4)\hspace{-0.049cm}+\hspace{-0.049cm}24411u^3
t^3(t^2\hspace{-0.049cm}+\hspace{-0.049cm}u^2)\hspace{-0.049cm}+\hspace{-0.049cm}34058u^4t^4){M}^4
\nonumber\\
&&\hspace{-0.4cm}+tu(334(t^7\hspace{-0.049cm}+\hspace{-0.049cm}u^7)\hspace{-0.049cm}-\hspace{-0.049cm}4775ut(t^5\hspace{-0.049cm}+\hspace{-0.049cm}u^5)\hspace{-0.049cm}-\hspace{-0.049cm}15750u^2t^2(t^3\hspace{-0.049cm}+\hspace{-0.049cm}u^3)\hspace{-0.049cm}-\hspace{-0.049cm}26417u^3t^3(t\hspace{-0.049cm}+\hspace{-0.049cm}u)){M}^2
\nonumber\\
&&\hspace{-0.4cm}+2t^2u^2(532(t^6\hspace{-0.049cm}+\hspace{-0.049cm}u^6)\hspace{-0.049cm}+\hspace{-0.049cm}1939ut(t^4\hspace{-0.049cm}+\hspace{-0.049cm}u^4)\hspace{-0.049cm}+\hspace{-0.049cm}3770u^2t^2(t^2\hspace{-0.049cm}+\hspace{-0.049cm}u^2)\hspace{-0.049cm}+\hspace{-0.049cm}4618u^3t^3))m^4
\hspace{-0.049cm}-\hspace{-0.049cm}tu(\hspace{-0.049cm}-\hspace{-0.049cm}4(7t^2
\nonumber\\
&&\hspace{-0.4cm}+33ut\hspace{-0.049cm}+\hspace{-0.049cm}7u^2){M}^{14}\hspace{-0.049cm}+\hspace{-0.049cm}(76(t^3\hspace{-0.049cm}+\hspace{-0.049cm}u^3)\hspace{-0.049cm}+\hspace{-0.049cm}844ut(t\hspace{-0.049cm}+\hspace{-0.049cm}u))
{M}^{12}\hspace{-0.049cm}+\hspace{-0.049cm}(188(t^4\hspace{-0.049cm}+\hspace{-0.049cm}u^4)\hspace{-0.049cm}-\hspace{-0.049cm}2303ut(t^2\hspace{-0.049cm}+\hspace{-0.049cm}u^2)
\nonumber\\
&&\hspace{-0.4cm}-3378u^2t^2)
{M}^{10}\hspace{-0.049cm}+\hspace{-0.049cm}(\hspace{-0.049cm}-\hspace{-0.049cm}718(t^5\hspace{-0.049cm}+\hspace{-0.049cm}u^5)\hspace{-0.049cm}+\hspace{-0.049cm}4033ut(t^3\hspace{-0.049cm}+\hspace{-0.049cm}u^3)\hspace{-0.049cm}+\hspace{-0.049cm}8973u^2t^2(t\hspace{-0.049cm}+\hspace{-0.049cm}u))
{M}^8\hspace{-0.049cm}+\hspace{-0.049cm}(762(t^6\hspace{-0.049cm}+\hspace{-0.049cm}u^6)
\nonumber\\
&&\hspace{-0.4cm}-5413ut(t^4\hspace{-0.049cm}+\hspace{-0.049cm}u^4)\hspace{-0.049cm}-\hspace{-0.049cm}15468u^2t^2(t^2\hspace{-0.049cm}+\hspace{-0.049cm}u^2)\hspace{-0.049cm}-\hspace{-0.049cm}21038u^3t^3){M}^6\hspace{-0.049cm}+\hspace{-0.049cm}(\hspace{-0.049cm}-\hspace{-0.049cm}334(t^7\hspace{-0.049cm}+\hspace{-0.049cm}u^7)\hspace{-0.049cm}+\hspace{-0.049cm}4775ut(t^5\hspace{-0.049cm}+\hspace{-0.049cm}u^5)
\nonumber\\
&&\hspace{-0.4cm}+15750u^2t^2(t^3\hspace{-0.049cm}+\hspace{-0.049cm}u^3)\hspace{-0.049cm}+\hspace{-0.049cm}26417u^3t^3(t\hspace{-0.049cm}+\hspace{-0.049cm}u)){M}^4\hspace{-0.049cm}+\hspace{-0.049cm}(54(t^8\hspace{-0.049cm}+\hspace{-0.049cm}u^8)\hspace{-0.049cm}-\hspace{-0.049cm}2182ut(t^6\hspace{-0.049cm}+\hspace{-0.049cm}u^6)
\hspace{-0.049cm}-\hspace{-0.049cm}8161u^2t^2(t^4\hspace{-0.049cm}+\hspace{-0.049cm}u^4)
\nonumber\\
&&\hspace{-0.4cm}-16101u^3t^3(t^2\hspace{-0.049cm}+\hspace{-0.049cm}u^2)\hspace{-0.049cm}-\hspace{-0.049cm}19812u^4t^4){M}^2\hspace{-0.049cm}+\hspace{-0.049cm}54tu(7(t^7\hspace{-0.049cm}+\hspace{-0.049cm}u^7)\hspace{-0.049cm}+\hspace{-0.049cm}30ut(t^5\hspace{-0.049cm}+\hspace{-0.049cm}u^5)
\hspace{-0.049cm}+\hspace{-0.049cm}68u^2t^2(t^3\hspace{-0.049cm}+\hspace{-0.049cm}u^3)\nonumber\\
&&\hspace{-0.4cm}+99u^3t^3(t\hspace{-0.049cm}+\hspace{-0.049cm}u)))m^2\hspace{-0.049cm}+\hspace{-0.049cm}2t^2u^2(\hspace{-0.049cm}-\hspace{-0.049cm}2{M}^2\hspace{-0.049cm}+\hspace{-0.049cm}t\hspace{-0.049cm}+\hspace{-0.049cm}u)^2(19
(t^2\hspace{-0.049cm}+\hspace{-0.049cm}ut\hspace{-0.049cm}+\hspace{-0.049cm}u^2){M}^8\hspace{-0.049cm}-\hspace{-0.049cm}(65(t^3\hspace{-0.049cm}+\hspace{-0.049cm}u^3)
\nonumber\\
&&\hspace{-0.4cm}+111ut(t\hspace{-0.049cm}+\hspace{-0.049cm}u))
{M}^6
\hspace{-0.049cm}+\hspace{-0.049cm}(100(t^4\hspace{-0.049cm}+\hspace{-0.049cm}u^4)\hspace{-0.049cm}+\hspace{-0.049cm}227ut(t^2\hspace{-0.049cm}+\hspace{-0.049cm}u^2)\hspace{-0.049cm}+\hspace{-0.049cm}300u^2t^2){M}^4\hspace{-0.049cm}-\hspace{-0.049cm}27(3(t^5\hspace{-0.049cm}+\hspace{-0.049cm}u^5)\hspace{-0.049cm}+\hspace{-0.049cm}8ut(t^3\hspace{-0.049cm}+\hspace{-0.049cm}u^3)
\nonumber\\
&&\hspace{-0.4cm}\hspace{-0.049cm}+\hspace{-0.049cm}13u^2t^2(t\hspace{-0.049cm}+\hspace{-0.049cm}u)){M}^2\hspace{-0.049cm}+\hspace{-0.049cm}27(t^2\hspace{-0.049cm}+\hspace{-0.049cm}ut\hspace{-0.049cm}+\hspace{-0.049cm}u^2)^3)\Big]
\end{eqnarray}


\providecommand{\href}[2]{#2}\begingroup\raggedright\endgroup

\end{document}